\documentclass[12pt]{article}
\pdfoutput=1
\usepackage{jheppub}
\usepackage{amsmath,amsfonts,braket,amssymb,bm,bbm,xcolor,graphicx}
\usepackage[small,labelfont=bf]{caption}
\usepackage{array}
\usepackage{relsize}
\usepackage{mathtools}
\usepackage{braket}

\usepackage{enumerate}
\usepackage{caption}
\usepackage{subcaption}
\usepackage{comment}

\allowdisplaybreaks
\newcommand\scalemath[2]{\scalebox{#1}{\mbox{\ensuremath{\displaystyle #2}}}}
\def\ov{\overline}

\usepackage[toc,page]{appendix}

\title{Differential equations for tree--level cosmological correlators with massive states}
\author[a]{Federico Gasparotto,}
\author[a]{Pouria Mazloumi,}
\author[a]{Xiaofeng Xu}

\affiliation[a]{PRISMA$^+$ Cluster of Excellence \& Mainz Institute for Theoretical Physics\\
    Johannes Gutenberg University, Staudingerweg 7, 55099 Mainz, Germany}

\emailAdd{fgasparo@uni-mainz.de}
\emailAdd{pmazloumi@uni-mainz.de}
\emailAdd{xiaxu@uni-mainz.de}

\abstract{We study mathematical aspects concerning two site tree-level cosmological correlators with massive internal and external states in a de Sitter universe. We employ integration by parts identities, (relative) twisted cohomology and the method of differential equations. We explicitly express the internally massive, externally conformally coupled correlator as a power series with respect to a small mass parameter, where the various terms in the series are given by multiple polylogarithms.}

\begin{document} 
\begin{flushright}
{\small
MITP-24-079\\
}
\end{flushright}

\maketitle

\newpage 
\section{Introduction}
In recent years there has been an increasing activity in studying cosmological correlators in de Sitter universe. The study of those correlators can give information concerning quantum fluctuations in the inflationary phase of the universe. Given the high energies accessible during inflation ($\approx10^{14}$ Gev) it is expected that both massless and massive modes would be created and participate in the quantum interactions of that phase. However only the massless fields are presumed to survive the inflationary phase while the massive modes leave an imprint in cosmological correlators. The study of those objects is part of the  research program known as ``Cosmological collider physics"~\cite{Arkani-Hamed:2015bza}. 

Several different new approaches have been developed, and new mathematical structures have been uncovered, in order to study those correlators. They include the method of differential equations, from the kinematic flow perspective~\cite{Arkani-Hamed:2023kig,Arkani-Hamed:2023bsv,Hang:2024xas,Baumann:2024mvm}, integration by parts identities or twisted cohomology~\cite{De:2023xue,Chen:2023iix,Benincasa:2024ptf}, their interplay with Gelfand--Kapranov--Zelevinsky (GKZ) systems or Euler Mellin integrals~\cite{Grimm:2024tbg,Fevola:2024nzj,He:2024olr}, the cosmological bootstrap program~\cite{Baumann:2022jpr,Arkani-Hamed:2018kmz}, cosmological optical theorem~\cite{Goodhew:2020hob}, cosmological CPT theorem~\cite{Goodhew:2024eup}, the connection with cosmological polytopes~\cite{Arkani-Hamed:2017fdk,Benincasa:2019vqr,Benincasa:2024leu}, Mellin-Barnes techniques~\cite{Sleight:2019mgd}, the method of regions~\cite{Beneke:2023wmt} and off-shell studies based on spectral representation~\cite{Werth:2024mjg}.

In particular, studies involving the method of differential equations are primarily focused on a conformally coupled scalar field, where there are advances both at tree-level and at loop-level and for general Friedmann–Robertson–Walker (FRW) cosmologies~\cite{Arkani-Hamed:2023kig,Arkani-Hamed:2023bsv,Hang:2024xas,Baumann:2024mvm,De:2023xue,Benincasa:2024ptf,He:2024olr}. Correlators are considered as members of an ``integral family", roughly speaking an infinite collection of integrals differing from the original one only for the powers of the factors appearing in the integrand. It is then possible to identify a basis for the above-mentioned infinite set of integrals, traditionally referred to as master integrals (MIs), and establish a system of differential equations fulfilled by the basis. Solving the system of differential equations we determine analytic expressions for the integral basis, thus circumventing the direct integration, which is often unfeasible. The ideas behind those methods have a long and well established tradition in the context of Feynman integrals and multi-loop calculus, going back to integration-by-parts-identities (IBPs)~\cite{Tkachov:1981wb,Chetyrkin:1981qh}, Laporta algorithm~\cite{Laporta:2000dsw}, and the method of differential equations~\cite{Kotikov:1990kg,Remiddi:1997ny,Gehrmann:1999as,Henn:2013pwa}--see e.g.~\cite{Weinzierl:2022eaz} for a comprehensive review. In recent years it was shown that Feynman integrals can be addressed from the perspective of twisted cohomology~\cite{Mastrolia:2018uzb,Frellesvig:2019uqt}, and, more precisely on a technical level, they can be studied borrowing techniques from relative twisted cohomology~\cite{Caron-Huot:2021xqj,Caron-Huot:2021iev}--see e.g.~\cite{aomoto2011theory,yoshida2013hypergeometric,Matsubara-Heo:2023ylc} for a textbook exposition on twisted cohomology. The framework of twisted (co)homology has found several applications in the context of Feynman integrals~\cite{Frellesvig:2019kgj,Frellesvig:2020qot,Mizera:2019vvs,Chen:2020uyk,Caron-Huot:2021xqj,Caron-Huot:2021iev,Chen:2022lzr,Cacciatori:2021nli,Giroux:2022wav,Abreu:2022mfk,Chestnov:2022alh,Chestnov:2022xsy,Jiang:2023oyq,Crisanti:2024onv,Duhr:2024rxe,Duhr:2024xsy,Fontana:2023amt,Brunello:2023rpq,Brunello:2024tqf,Chen:2024ovh}, as well as in different branches of theoretical physics~\cite{Mizera:2017cqs,Mizera:2017rqa,Casali:2019ihm,Mizera:2019gea,Weinzierl:2020nhw,Mazloumi:2022nvi,Gasparotto:2022mmp,Cacciatori:2022mbi,Gasparotto:2023roh,Brunello:2023fef,Bhardwaj:2023vvm,Frellesvig:2024swj,Mazloumi:2024wys}. 

In this work we study tree-level cosmological correlators with massive internal and external states, using (relative) twisted cohomology, as well as IBP identities, and the method of differential equations. Massive modes are given in terms of Hankel functions; this fact makes the calculation of the correlators more involved compared to the conformally coupled case.

This manuscript is organized as follows; in Section \ref{sec2} we give a short and concise introduction to the cosmological model that we are using, in particular the Lagrangian for a massive (and conformally coupled massless) scalar in de Sitter universe, solutions to the equations of motions as well as the corresponding Green functions. We arrive at an integral representation for the correlators of our interest consisting of products of hyperplanes. Section \ref{sec3} is the technical part of this work; we introduce (relative) twisted cohomology and use some of the tools in this framework; in particular we exploit the existence of a certain ``dual space" which is isomorphic to the original space and more convenient for calculations, to derive the system of differential equations. We also propose an alternative, yet equivalent, strategy to obtain the above-mentioned system, without relying on the technicalities of (relative) twisted cohomology. In Section \ref{sec4} we discuss the solution of the differential equations as an expansion with respect to a parameter $\delta$ which is associated to a small mass. In Section \ref{sec5} we propose some observations concerning the structure of the singularities of the integrals we are studying and a simple algorithm which, at least for the setup we are considering, gives us the possibility to predict all the singularities appearing in the differential equations. In particular they can be understood as the Pl\"ucker coordinates associated to the hyperplanes appearing in the integral representation for the correlators.\\
~\\
Through this work we employed \textsc{xAct}~\cite{xAct} for differential algebra,  \textsc{MultivariateApart}~\cite{Heller:2021qkz} for partial fraction decomposition, \textsc{PolyLogTools}~\cite{Duhr:2019tlz} for handling multiple polylogairhtms and \textsc{GiNaC}~\cite{Vollinga:2004sn} for their numerical evaluation, and \textsc{HypExp2}~\cite{Huber:2007dx} for the  expansion of hypergeometric functions.\\
~\\
\textbf{Note added: }while this manuscript was in preparation Refs.~\cite{Fevola:2024nzj,Chen:2024glu} appeared. In Ref.~\cite{Fevola:2024nzj} the authors describe the singularities appearing in the differential equation in terms of the Euler discriminant, and its defining polynomial, built upon a certain matrix associated to the hyperplane arrangements defining the integral. Our findings in Section \ref{sec5} are similar. In Ref.~\cite{Chen:2024glu} differential equations for cosmological correlators involving massive propagators are discussed, and analytic powers series solutions are obtained. Our mathematical set-up for tackling the problem is different.

\section{Notations and Setup} \label{sec2}

The main goal of this Section is to review basic ideas and ingredients in order to arrive at the integral representation for the correlator of our interest. Far from being a detailed and a comprehensive treatment, we follow here a pragmatic approach (for a detailed discussion see e.g. \cite{Benincasa:2022gtd}).

Let us consider the following Lagrangian
\begin{equation}
\begin{aligned}
S= \int d^{d}x  \,\,\int_{-\infty}^0 d\eta \, \sqrt{-g}  \Big[ & \frac{1}{2} g^{\mu \nu} \partial_{\mu} \sigma \partial_{\nu} \sigma-\frac{1}{2}\xi R \sigma^2+\frac{1}{2} g^{\mu \nu} \partial_{\mu} \Phi \partial_{\nu} \Phi  
\\
-&\frac{1}{2} \left(m^2+\xi R \right) \Phi^2 - \mathcal{L}_{\text{int}}\Big],
\label{eq:Lagrangian1}
\end{aligned}
\end{equation}
where $R$ is the Ricci scalar, $\eta$ is referred to as conformal time and $d$ denotes the number of spatial dimensions. The most generic polynomial interaction among the fields $\Phi$ and $\sigma$ is
 \begin{equation}
     \mathcal{L}_{\text{int}}= \sum_{m=3}^{\infty}\mathcal{L}_{m}, \qquad \mathcal{L}_{m}= \sum_{k=0}^{m} \frac{\lambda_{m-k,k}}{(m-k)! k!} \Phi^{m-k} \sigma^{k}.
 \label{eq:L_int_generic}
 \end{equation}
 
The metric for Friedmann–Robertson–Walker (FRW) cosmology is defined as
\begin{align}
ds^2=a(\eta)^2 (- d \eta^2 + dx^i dx_i)\,, 
\end{align}
where the index $i$ runs over the spatial components $i=1,2,\dots,d$. The scale factor is given by $a(\eta) \propto (-\eta)^{-(1+\varepsilon)}$, and different values of $\varepsilon$ corresponds to different cosmological scenarios. In this work we restrict to the case $\varepsilon=0$ corresponding to a de-Sitter cosmology. The Ricci scalar $R$ in the action with the FRW metric reads 
\begin{align}
R=\frac{2d}{a^2} \left[\partial_{\eta}\left( \frac{\Dot{a}}{a}\right)+\frac{d-1}{2} \left( \frac{\Dot{a}}{a}\right)^2\right]\,.
\end{align}
With the conformal transformations
\begin{align}
g_{\mu \nu} \to a(\eta)^2 g_{\mu \nu} \, , \quad \sigma \to a(\eta)^{ -\frac{d-1}{2}} \sigma \, , \quad \Phi \to a(\eta)^ {-\frac{d-1}{2}} \Phi \, ,
\end{align}
the action can be rewritten as
\begin{equation}
S= \int d^{d}x  \,\,\int_{-\infty}^0 d\eta \, \Big[\frac{1}{2} (\partial \sigma)^2 +\frac{1}{2} (\partial \Phi)^2-\frac{1}{2} \mu_\sigma^2(\eta) \sigma^2-\frac{1}{2} \mu_\Phi^2(\eta) \Phi^2-\mathcal{L}_{\text{int}}(\eta) \Big]\,.
\label{eq:Lagrangian}
\end{equation}
The interaction term $\mathcal{L}_{\text{int}}(\eta)$ is of the same form as eq.~(\ref{eq:L_int_generic}), but with scaling factor dependent couplings, i.e. it is obtained upon the substitution 
\cite{Arkani-Hamed:2023kig,Hillman:2019wgh}:
\begin{equation}
\begin{aligned}
    & \lambda_{m-k,k} \to \lambda_{m-k,k}(\eta)= \lambda_{m-k,k}[a(\eta)]^{2+\frac{(d-1)(2-m)}{2}}\,. 
\end{aligned}
\end{equation}
$\mu_\sigma$ and $\mu_\Phi$ corresponds to effective masses for the fields $\sigma$ and $\Phi$; the explicit expressions read
\begin{equation}
\begin{aligned}
& \mu_\Phi^2(\eta)=a^2(\eta) m^2 + a^2(\eta)\left(\xi- \frac{d-1}{4d} \right) R \,, \\
&\mu_\sigma^2(\eta)= a^2(\eta)\left(\xi- \frac{d-1}{4d} \right) R \,.
\end{aligned}
\end{equation}
Two cases are particularly relevant: the case $\xi=\frac{d-1}{4d}$, referred to as conformal coupling\footnote{This means that the mass term in this choice for $\xi$ is invariant under the rescaling of fields.}, and the case $\xi=0$, known as minimal coupling. In this paper, we will focus on the conformal coupling, and restrict ourselves to cubic interaction, i.e. $\mathcal{L}_{\text{int}}=\mathcal{L}_3$. The action can be simplified as
\begin{equation}
S= \int d^{d}x  \,\,\int_{-\infty}^0 d\eta \, \Big[-\frac{1}{2} (\partial \sigma)^2 -\frac{1}{2} (\partial \Phi)^2-\frac{1}{2} \mu^2(\eta) \Phi^2-\mathcal{L}_3(\eta)  \Big]\,,
\label{eq:Lagrangian}
\end{equation} 
with the mass $\mu^2(\eta)=m^2 a^2(\eta)=m^2(-\eta)^{-2}$ (where the last equality holds in de Sitter).
~\\

\begin{figure}[h]
\begin{subfigure}{1.\linewidth}
\centering
\includegraphics[scale=.30]{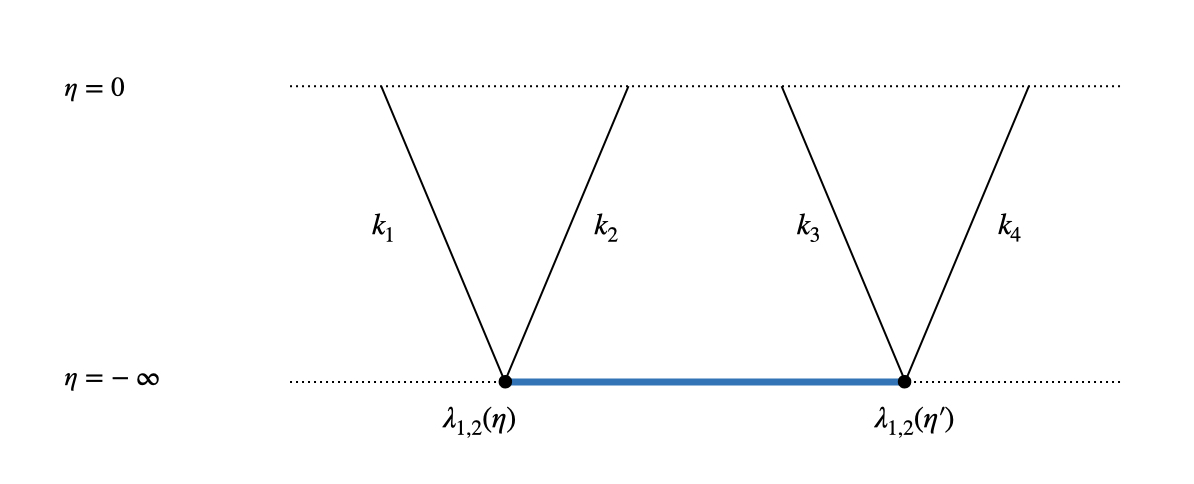}
\caption{Two site tree-level correlator with massive internal state and conformally coupled external states.}
\label{Fig:massless}
\end{subfigure}

\begin{subfigure}{1.\linewidth}
\centering
\includegraphics[scale=.30]{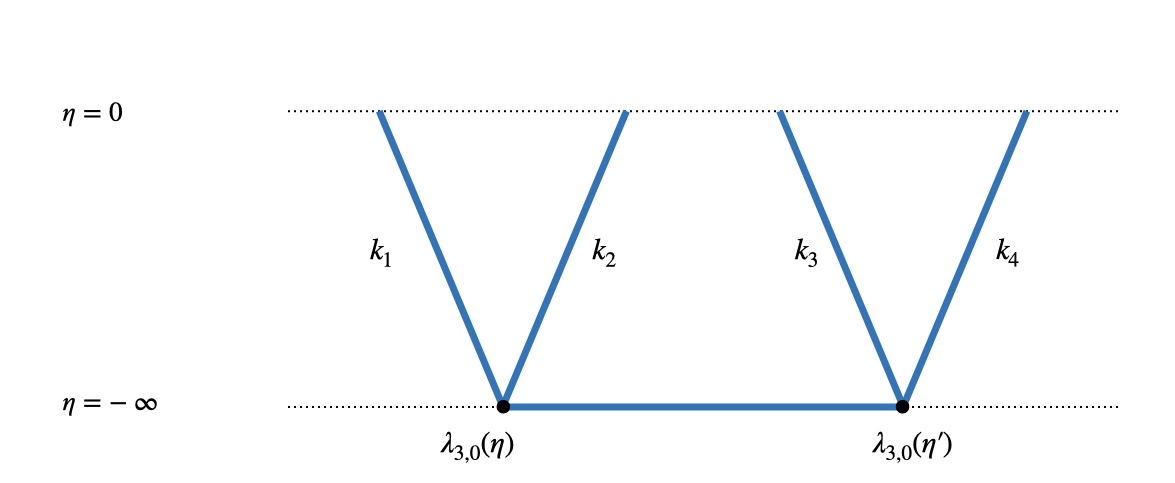}
\caption{Two site tree-levels correlator with massive internal state and massive external states.}
\label{Fig:massive}
\end{subfigure}
\end{figure}

The mode functions for the fields $\sigma(\eta)$ and $\Phi(\eta)$ are determined by the following differential equations
\begin{equation}
\begin{split}
    &  \sigma''_0(k,\eta)+k^2\sigma_{0}(k,\eta)=0\,,\\
    & \Phi''_0(k,\eta)+(k^2+\mu^2(\eta))\Phi_{0}(k,\eta)=0, \quad\quad \text{with: } k= |\vec{k}|\,.
\label{eq:EOM_phi}
\end{split}
\end{equation}
Explicit solutions for eq.~(\ref{eq:EOM_phi}) are given by \cite{Chen:2023iix,Benincasa:2024leu} 
\begin{equation}
\begin{split}
      \sigma_0(k,\eta)&=e^{i k \eta}\,,\\
     \Phi^{\nu}_0(k,\eta)&=-i \sqrt{\frac{\pi}{2}}\sqrt{-k\eta} \, H_{\nu}^{(2)}(-k\eta)\,,\\ 
\end{split}
\label{eq:EOM_explicit_Bulk_to_Boundary}
\end{equation}
where $H^{(2)}_{\nu}$ denotes the Hankel function of the second type, and $\nu=\sqrt{\frac{1}{4}-m^2}$. We notice that $\nu$ can be real (for small mass) or purely imaginary (for heavy mass). In rest of this work we assume $\nu$ to be real. The limit $m^2 \to 0$ corresponds to $\nu \to \frac{1}{2}$; with the explicit normalisation in eq.~(\ref{eq:EOM_explicit_Bulk_to_Boundary}) we have
\begin{equation}
    \lim_{\nu \to \frac{1}{2}}\Phi^{\nu}_0(\eta)=e^{i k \eta}\,.
\label{eq:limit_Hankel}
\end{equation}
We should point out that, in order for the mode function in eq.~(\ref{eq:EOM_explicit_Bulk_to_Boundary}) to decay at $\eta \rightarrow -\infty$, it is necessary to assign a small imaginary part to $\eta$ i.e. $\eta \to \eta-i \eta \,0^+ $. \\
In this work we are interested in two site tree-level correlators with massive internal states and either conformally coupled external states or massive external states. The corresponding graphs are shown in Figure~\ref{Fig:massless} and Figure~\ref{Fig:massive}.\\
~\\
In the case of conformally coupled external states the correlator is given by
\begin{equation}
\begin{split}
\psi_2(k_{12},k_{34},Y)=  \int_{-\infty}^0  d^2 \eta\, & (i \lambda_{1,2}(\eta)) \sigma_0(k_1,\eta) \sigma_0(k_2,\eta) \times G(Y, \eta, \eta') \times \\ 
& (i \lambda_{1,2}(\eta')) \sigma_0(k_3,\eta')\sigma_0(k_4,\eta')\,,
\end{split}
\label{eq:correlator_ext_massless}
\end{equation}
where $k_{ij}=k_i+k_j$. The correlator for external massive states is obtained from eq.~(\ref{eq:correlator_ext_massless}), upon the replacements $\sigma_0 \to \Phi^{\nu}_0$ and $\lambda_{1,2} \to \lambda_{3,0}$.\\
~\\
The quantity $G(Y,\eta,\eta')$ in eq.~(\ref{eq:correlator_ext_massless}) is referred to as the bulk to bulk propagator of the massive field $\Phi$ \cite{Goodhew:2021oqg}: 

\begin{equation}
\begin{aligned}
G(Y,\eta,\eta')=\frac{1}{2 Y}\Big[&\overline{\Phi^\nu_0}(-Y,\eta) \Phi^\nu_0(-Y,\eta')\theta(\eta-\eta')+\Phi^\nu_0(-Y,\eta) \overline{\Phi^\nu_0}(-Y,\eta')\theta(\eta'-\eta)\\
& +h_0\Phi^{\nu}_0(-Y,\eta) \Phi^{\nu}_0(-Y,\eta')\Big]\,,
\label{eq:ebulkprop}
\end{aligned}
\end{equation}
where $\overline{\Phi^\nu_0}$ denotes the complex conjugate of $\Phi^\nu_0$, and\footnote{Under the assumption $\nu$ and $Y$ real and positive.} $h_0 \equiv -\lim_{\eta \to 0^-} \frac{\overline{\Phi^{\nu}_0}(-Y, \eta)}{\Phi^{\nu}_0(-Y, \eta)}=-1$. Considering explicit expressions given in eq.~(\ref{eq:EOM_explicit_Bulk_to_Boundary}), we obtain
\begin{equation}
\begin{aligned}
G(Y,\eta,\eta')=
\frac{\pi }{4}\sqrt{-\eta}\sqrt{- \eta'}\Big[& \ov{H^{(2)}_\nu}(-Y\eta) H^{(2)}_\nu(-Y\eta')\theta(\eta-\eta')+H^{(2)}_\nu(-Y\eta) \ov{H^{(2)}_\nu}(-Y\eta')\theta(\eta'-\eta)\\
& -h_0{H^{(2)}_\nu}(-Y\eta) H^{(2)}_\nu(-Y\eta')\Big]\,. \label{Green}
\end{aligned}
\end{equation}

\subsection{Integral representation for correlators with massive propagator} \label{sec:correlatot_def}
In order to proceed with the discussion, we prefer to express the Hankel functions associated to massive states in terms of the modified Bessel function $K_{\nu}$. We consider the relations \cite{abramowitz+stegun}
\begin{equation}
\begin{aligned}
&\overline{H_\nu^{(2)}(z)}=H_{\overline{\nu}}^{(1)}(\overline {z}) \, ,
 \\
& H_{{-\nu}}^{(1)}({z})=e^{i \pi  \nu} H_{{\nu}}^{(1)}({z})\, ,\\
&H_{{\nu}}^{(1)}({z})= -\frac{(2 i)}{\pi} \exp \left(-\frac{1}{2} i \pi  \nu \right) K_{\nu }(-i z)  \, , \quad  -\frac{\pi}{2} < \arg(z) < \pi \, ,
\\
&H_{{\nu}}^{(2)}({z})= \frac{(2 i)}{\pi} \exp \left(\frac{1}{2} i \pi  \nu \right) K_{\nu }(i z)  \, ,  \quad \quad \quad\,\,\, -\pi  < \arg(z) < \frac{\pi}{2}\,.
\label{eq:Hankel_relation}
\end{aligned}
\end{equation}
The argument of the Hankel functions, denoted in eq.~(\ref{eq:Hankel_relation}) via $z$, has a small imaginary part inherited from $\eta$. 

With the relation eq.~(\ref{eq:Hankel_relation}),  the bulk to bulk propagators in eq.~(\ref{eq:ebulkprop}) can be rewritten as
\begin{equation}
\begin{aligned}
G(Y,\eta,\eta')= &\frac{1}{\pi}\sqrt{-\eta} \sqrt{-\eta'}\Big[K_{\nu}(iY\ov{\eta}) K_\nu(-iY\eta')\theta(\eta-\eta')+K_\nu(-iY\eta) K_{\nu}(iY\ov{\eta'})\theta(\eta'-\eta)\\ 
& +e^{i \pi \nu} h_0 K_{{\nu}}({-iY\eta}) K_\nu(-iY\eta')\Big] \, .
\label{eq:prop_K}
\end{aligned}
\end{equation}
The Bessel function $K_{\nu}$ admits the integral representation \cite{Bessel_Rep}
\begin{equation}
K_{\nu}(w)=\frac{\pi^{1/2} (\frac{1}{2}w)^{\nu}}{\Gamma(\nu+\frac{1}{2})}\int_{1}^\infty dt \, e^{-wt} (t^2-1)^{\nu-\frac{1}{2}}\,, \quad \quad {\small \Re(\nu)>-\frac{1}{2}\,,\,\, \Re (w)>0}\,.
\label{eq:Bessel_K}
\end{equation}
The condition $\Re (w)>0$ is fulfilled due to the imaginary part assigned to $\eta$. A notable feature of this integral representation is the simplicity of the expression within the exponential. In addition, the factor $t^2-1$ can be factorized in terms of  two hyperplanes: $t-1$ and $t+1$.\\ 
~\\
Now we investigate the two-site correlator with conformally coupled external states eq.~(\ref{eq:correlator_ext_massless}) in $d=3-2\epsilon$  (see Figure~\ref{Fig:massless}). Using the massive bulk to bulk propagator eq.~(\ref{eq:prop_K}) and the integral representation of eq.~(\ref{eq:Bessel_K}), the correlator can be rewritten as\footnote{For ease of notation we omit writing explicitly the small imaginary part of $\eta$.}:
\begin{equation}
\begin{split}
\psi_2(k_{12},k_{34},Y)=&  \left( \frac{ i \lambda_{1,2}   Y^\nu  }{ 2^{\nu } \Gamma \left(\nu +\frac{1}{2}\right) }  \right)^2  \times \\
& \,\int_{-\infty}^0 d^2\eta \, (-\eta)^{-\frac{1}{2}-\epsilon+\nu}\, (-\eta^\prime)^{-\frac{1}{2}-\epsilon+\nu} \int_1^{\infty} d^2 t \, (t_1^2-1)^{\nu-1/2}(t_2^2-1)^{\nu-1/2} \times \\ 
&  \exp[ik_{12}\eta+ik_{34}\eta^\prime] \, \Big[ \exp[-i Y\left( \eta t_1-\eta^\prime t_2\right)] \theta \left( \eta -\eta^\prime \right)+ \\
& \qquad \qquad \qquad \quad \,\,\,\,\,\, \,\,\,\, \exp[-i Y\left( \eta^\prime t_2 -\eta t_1 \right)] \theta \left( \eta^\prime-\eta\right)+\\
& \qquad \qquad \qquad \quad \,\,\,\,\,\, \,\,\,\, h_0 \exp( 2 i \pi \nu)  \exp[i Y\left( \eta t_1 +\eta^\prime t_2\right)]  \Big] \, .
\end{split}
\end{equation}
We choose to work in $d=3-2\epsilon$ (and not $d=3$) in order to regulate otherwise divergent terms appearing in intermediate stages of the calculations\footnote{We will be more precise about this point later.}. 
\\

Further, we consider the Fourier transform of the factor $ (-\eta)^{-\frac{1}{2}-\epsilon+\nu} $ in the integrals, according to 
\begin{align}
(-\eta)^{-\frac{1}{2}-\epsilon+\nu}= 
 \frac{\exp\left(-\frac{1}{2} i \pi  \left(\nu -\epsilon -\frac{1}{2}\right) \right) }{\Gamma \left(\epsilon -\nu +\frac{1}{2}\right)} \int_{0}^\infty d x  x^{-\frac{1}{2}+\epsilon-\nu} e^{-i(-\eta)x}\,; 
\label{eq:Fourier_transform}
\end{align}
notice that the integral in eq.~(\ref{eq:Fourier_transform}) is convergent at infinity thanks to the prescription $\eta \to \eta-i \eta \,0^+$ mentioned above. \\
Finally, we  can integrate out the variables $\eta$ and $\eta^\prime$, and obtain the integral representation 
\begin{equation}
\begin{aligned}
\psi_2(k_{12},k_{34},Y) =& \,c_1\, Y^{2 \nu} \prod_{i=1}^2 \int_{0}^{\infty} dx_i x_i ^{-\frac{1}{2}+\epsilon-\nu}     \prod_{j=1}^2 \int_1^{\infty}d t_j   (t_j^2-1)^{\nu-1/2}  
\\ 
&\Big[\frac{1}{\left(k_{34}+t_2 Y+x_2\right) \left(k_{12}+k_{34}-t_1 Y+t_2 Y+x_1+x_2\right)}\, 
\\
&+ \frac{1}{\left(k_{12}+t_1 Y+x_1\right) \left(k_{12}+k_{34}+t_1 Y-t_2 Y+x_1+x_2\right)}
\\
&+\frac{ \exp(2 i \pi \nu)  h_0 }{\left(k_{12}+t_1 Y+x_1\right) \left(k_{34}+t_2 Y+x_2\right)} \Big] \, ,  \label{eq:correlator_complete}
\end{aligned}
\end{equation}
where the constant $c_1$ is given by
\begin{equation}
   c_1= \left(\frac{  i \lambda_{1,2}  }{ 2^{\nu} \Gamma \left(\nu +\frac{1}{2}\right) \Gamma \left(\epsilon -\nu +\frac{1}{2}\right)}\right)^2  \exp \left(i \pi  ( \epsilon -\nu-\frac{1}{2}) \right)\,.
\end{equation}
All the factors in the denominators in eq.~(\ref{eq:correlator_complete}) carry a negative imaginary part $-i0^{+}$ (once again inherited from $\eta$) which, from now on, we give as understood.

The correlator in eq.~(\ref{eq:correlator_complete}) consists of three terms, dubbed as right-center ($\text{RC}$), left-center ($\text{LC}'$), and left-right ($\text{LR}$) according to Figure~\ref{Fig:massless}\footnote{This labeling is related to the kinematic dependence of the differential forms appearing in the integral. }.\\
The right-center contribution is given by
\begin{equation}
    \begin{split}
        \psi^{RC}_2(k_{12},k_{34},Y)=&\,c_1 \, Y^{2 \nu} \prod_{i=1}^2 \int_{0}^{\infty} dx_i x_i ^{-\frac{1}{2}+\epsilon-\nu}     \prod_{j=1}^2 \int_1^{\infty}d t_j   (t_j^2-1)^{\nu-1/2} \times \\
        & \frac{1}{\left(k_{34}+t_2 Y+x_2\right) \left(k_{12}+k_{34}-t_1 Y+t_2 Y+x_1+x_2\right)}\,,  
    \end{split}
    \label{eq:psi_RC}
\end{equation}
the left-center\footnote{Notice that the term $k_{12}{+}k_{34}{-}t_1 Y{+}t_2 Y{+}x_1{+}x_2$ in eq.~(\ref{eq:psi_RC}) is different from the term  $k_{12}{+}k_{34}{+}t_1 Y{-}t_2 Y{+}x_1{+}x_2$ appearing in eq.~(\ref{eq:psi_LC}), hence we use the labelling $\psi_2^{LC'}$ (and not $\psi_2^{LC}$).} reads
\begin{equation}
    \begin{split}
        \psi^{LC'}_2(k_{12},k_{34},Y)=&\, c_1 \, Y^{2 \nu} \prod_{i=1}^2 \int_{0}^{\infty} dx_i x_i ^{-\frac{1}{2}+\epsilon-\nu}     \prod_{j=1}^2 \int_1^{\infty}d t_j   (t_j^2-1)^{\nu-1/2} \times \\
        & \frac{1}{\left(k_{12}+t_1 Y+x_1\right) \left(k_{12}+k_{34}+t_1 Y-t_2 Y+x_1+x_2\right)}\,,  
    \end{split}
    \label{eq:psi_LC}
\end{equation}
and finally the left-right is
\begin{equation}
\begin{split}
    \psi^{LR}_2(k_{12},k_{34},Y)=&\,c_1\, Y^{2 \nu} \prod_{i=1}^2 \int_{0}^{\infty} dx_i x_i ^{-\frac{1}{2}+\epsilon-\nu}     \prod_{j=1}^2 \int_1^{\infty}d t_j   (t_j^2-1)^{\nu-1/2} \times \\
    & \frac{ \exp(2 i \pi \nu) h_0 }{\left(k_{12}+t_1 Y+x_1\right) \left(k_{34}+t_2 Y+x_2\right)}\,.
\end{split}
\label{eq:psi_LR}
\end{equation}
The integrands are written in terms of hyperplanes in the variables $(t_1, t_2, x_1, x_2)$. Some of the hyperplanes are raised to a non-integer power, given the presence of the parameter $\nu$ and the regulator $\epsilon$ and thus give a multivalued term in the integrand. The calculations of the integrals in eqs.~(\ref{eq:psi_RC}),\,(\ref{eq:psi_LC}) and~(\ref{eq:psi_LR}) is one of the main goal of this work. In this respect, we notice that eqs.~(\ref{eq:psi_RC}) and~(\ref{eq:psi_LC}) are related via
\begin{equation}
\psi_2^{LC'}(k_{12},k_{34},Y) = \psi_2^{RC}(k_{12},k_{34},Y) \big|_{k_{12}\leftrightarrows k_{34}}.
\end{equation}
The integral in eq.~(\ref{eq:psi_LR}) can be obtained by direct integration; therefore the only non trivial task is to derive an integrated expression for eq.~(\ref{eq:psi_RC}).\\
The fully massive case involves products of six Hankel functions; hence we are led to a eight fold integral representation 
\begin{equation}
\begin{aligned}
\psi_{2,\text{mass}} = & c_2 Y^{2 \nu} \prod_{j=1}^4 k_j ^{1/2+\nu} \prod_{i=1}^2 \int_{0}^{\infty} dx_i x_i ^{-3 \nu +\epsilon -\frac{3}{2}}   \prod_{l=1}^6 \int_1^{\infty}d t_l   (t_l^2-1)^{\nu-1/2}  \times
\\
& \Big[ \frac{1}{\left(k_3 t_3{+}k_4 t_4{+}t_6 Y{+}x_2\right) \left(k_1 t_1{+}k_2 t_2{+}k_3 t_3{+}k_4 t_4{-}t_5 Y{+}t_6 Y{+}x_1{+}x_2\right)}
\\
&+\frac{1}{\left(k_1 t_1{+}k_2 t_2{+}t_5 Y{+}x_1\right) \left(k_1 t_1{+}k_2 t_2{+}k_3 t_3{+}k_4 t_4{+}t_5 Y{-}t_6 Y{+}x_1{+}x_2\right)} 
\\
&+\frac{h_0  \exp(2 i \pi \nu)}{\left(k_1 t_1{+}k_2 t_2{+}t_5 Y{+}x_1\right) \left(k_3 t_3{+}k_4 t_4{+}t_6 Y{+}x_2\right)} \Big] \, ,
\label{eq:psi_2_massive}
\end{aligned} 
\end{equation}
where the constant $c_2$ is
\begin{align}
c_2=\frac{ \exp \left(i \pi  (\nu +\epsilon +\frac{1}{2} ) \right) }{ \left(2^ \nu \Gamma \left(\nu +\frac{1}{2}\right) \right)^6}  \left(\frac{ 2 i \lambda_{3,0} }{ \Gamma \left(\epsilon -3 \nu -\frac{1}{2}\right)}\right)^2 \, .
\end{align}
We notice that the integral in eq.~(\ref{eq:psi_2_massive}) exhibits the same decomposition as eq.~(\ref{eq:correlator_complete}); it is the sum of three contributions: a right-center term (dubbed $\psi^{RC}_{2,\text{mass}}$),  a left-center term (dubbed $\psi^{LC'}_{2,\text{mass}}$)  and left-right term (dubbed $\psi^{LR}_{2,\text{mass}}$). They correspond to the first, second and third line in eq.~(\ref{eq:psi_2_massive}) respectively.
\section{Construction of differential equations } \label{sec3}

\subsection{A short review of twisted cohomology}
The integral representation of cosmological correlators described above fall into the category of Aomoto-Gelfand hypergeometric integrals, 
\begin{align}
I(\varphi)=\int_{\mathcal{C}} u  \varphi\,, \qquad \qquad u = \prod_{i=1}^m l_i^{\alpha_i}\,,  \qquad \qquad \alpha_i  \in \mathbb{C} \setminus \mathbb{Z}\,,
\label{eq:hyperint}
\end{align}
where $l_i$ defines an hypersurface in $\mathbb{C}^n$, $\varphi$ is an holomorphic $(n,0)$-form on manifold $X=\mathbb{C}^n\setminus( \cup^m_{i=1} \, l_i=0)$. We should notice that the system in eq.~(\ref{eq:hyperint}) is originally defined over the complex projective space $\mathbb{CP}^n$; however in this section, as well as in the rest of this work, we always assumes a specific chart $\mathbb{C}^n$ of $\mathbb{CP}^n$. Furthermore we assume the following condition
\begin{equation}
u\big|_{(\partial \mathcal{C})}=0.    
\label{eq:u_vanishing_boundary}
\end{equation}
The space of holomorphic differential $(n,0)$-forms is denoted as $\Omega^n(X, \nabla_{\omega})$\,.\\
~\\
It is important to note that, in general, integrals and hence differential forms, are not linearly independent, thanks to integration by parts identities. Let us consider an $(n-1)$ differential form dubbed as $\xi$; then using  eq.~(\ref{eq:u_vanishing_boundary}) Stokes' theorem gives 
\begin{align}
0= \int_{\mathcal{C}}d (u \xi)=\int_{\mathcal{C}} u(d+\omega\wedge) \xi \equiv \int_{\mathcal{C}} u\nabla_{\omega} \xi\,,
\label{eq:IBP_twisted_cohomolohy}
\end{align}
where we introduced the connection
\begin{equation}
     \nabla_{\omega}(\bullet) = d(\bullet)+ \omega \wedge \bullet\,, \qquad \text{with: $\omega=d \log u$}\,.
\label{eq:nabla_omega}
\end{equation}
Utilizing eq.~(\ref{eq:IBP_twisted_cohomolohy}), we can study differential forms in the framework of twisted cohomology. In a nutshell, within this framework we do not consider differential forms separately, rather we group them into equivalence classes. Those equivalence classes, dubbed as twisted cocycles, are element of the $n$th twisted cohomology group\footnote{In \cite{aomoto1975vanishing} it was shown that all the $\operatorname{H}^m$, $m \neq n$ are empty.}  

\begin{equation}
    \operatorname{H}^n(X,\nabla_{\omega})=\frac{\{\varphi \in \Omega^n(X, \nabla_{\omega}) \, | \,  \nabla_{\omega} \varphi=0 \}}{\{\phi \in \Omega^n(X, \nabla_{\omega}) \,| \, \phi=\nabla_{\omega}\xi \}}\,.
\label{eq:cohomology_group}
\end{equation}

In words the $n$th cohomology group is defined as the quotient of the  closed $n$-from by exact $n$-forms, under the action of the covariant derivative defined in eq.~(\ref{eq:nabla_omega}).\\
Integrals as the ones in eq.~(\ref{eq:hyperint}) are regarded as the pairing of a twisted cocycle and a certain twisted cycle (loosely speaking containing the information about the topological cycle $\mathcal{C}$ with a branch of the multivalued function $u$ on it). Studying the $n$th cohomology group eq.~(\ref{eq:cohomology_group}), we can infer information for the integrals in eq.~(\ref{eq:hyperint}).\\
In particular, the $n$th cohomology group has finite dimension; the latter corresponds to the number of independent integrals modulo integration by parts. As a consequence any integral of the form of eq.~(\ref{eq:hyperint}) can be expressed as a linear combination of finitely many independent integrals, known as master integrals, associated to the basis of the cohomology. The problem of evaluating any integral in eq.~(\ref{eq:hyperint}), boils down to the evaluation of the master integrals. In this work we denote the $i-$th element in the basis of the cohomology as $\phi_i$ and the corresponding master integrals as $\text{I}(\phi_i)$.\\
~\\
Under mild assumptions we have the following
\begin{equation}
    \dim \operatorname{H}^n(X, \nabla_{\omega}) = \text{Number of Critical Points of $ \log u$}\,, \label{eq:dimhom}
\end{equation}
where a critical point $z_{\text{crt}}$ for $u$ is defined as
\begin{equation}
z_{\text{crt}} \in \mathbb{C}^n \, \,  | \, \, \omega|_{z=z_{\text{crt}}} = d \log u|_{z=z_{\text{crt}}}=0\,.     \end{equation}
From a more abstract and topological view point, we have that the dimension of the cohomology group amounts the number of \emph{bounded chambers} enclosed by the hyperplanes defining $u$ (cf. eq.~(\ref{eq:hyperint})). Through this work we always assume this criterium, as well as eq.~(\ref{eq:dimhom}), to be valid.\\
~\\
In this work, we exploit the differential equations method to compute the master integrals. Therefore, we need to define the derivative of the master integrals with respect to kinematic variables; we have
\begin{align}
d_{\text{kin}} \text{I}(\phi_i)=\int_{\mathcal{C}} u \left( d_{\text{kin}}\phi_i+ d_{\text{kin} } \log u \wedge \phi_i \right)=\int_{\mathcal{C}} u \left( d_{\text{kin}}\phi_i+ \omega_{\text{kin}} \wedge \phi_i \right)= \int_{\mathcal{C}} u \nabla_{\text{kin}} \phi_i \,,
\label{eq:nabla_kin_op}
\end{align} 
where we introduced the short-hand notation
\begin{equation}
    d_{\text{kin}}(\bullet) = d k_{12} \,  \frac{\partial(\bullet)}{\partial k_{12}} + d k_{34}  \frac{\partial(\bullet)}{\partial k_{34}} + dY \frac{\partial(\bullet)}{\partial Y}\,, \qquad \omega_{\text{kin}}= d_{\text{kin}} \log u\,.
\end{equation}
So there is a correspondence between the derivative of an  integral and the covariant derivative of the corresponding differential form, i.e. we have
\begin{equation}
    d_{\text{kin}} \text{I} (\phi_i) = I \left( \nabla_{\text{kin}} \phi_i \right)\,.
\label{eq:nabla_kin_correspondence}
\end{equation}
Expressing the right hand side of eq.~(\ref{eq:nabla_kin_correspondence}) as a linear combination of master integrals, and repeating the same procedure for all the elements in the basis, we obtain a system of first order differential equations denoted as
\begin{equation}
    d_{\text{kin}} \mathbf{I}(k_{12}, k_{34}, Y) =\mathbf{A}(k_{12}, k_{34}, Y) \, \, \mathbf{I}(k_{12}, k_{34}, Y)\,,
\label{eq:DEQ_generic}
\end{equation}
with (omitting the dependencies on $k_{12}, k_{34}$ and $Y$)
\begin{equation}
    \mathbf{A}=\mathbf{A}_{k_{12}} d k_{12}+\mathbf{A}_{k_{34}} d k_{34}+\mathbf{A}_{Y} d Y\,.
\end{equation}
The solution of eq.~(\ref{eq:DEQ_generic}), supplemented with an appropriate set of boundary conditions, enables the determination of the master integrals. 
\subsection{Twisted cohomology of internally massive correlator with conformally coupled external states }
Let us focus on the integrals appearing in the study of eq.~(\ref{eq:correlator_complete}). In order to consider them from the twisted cohomology perspective (cf. eq.~(\ref{eq:hyperint})), we need the following identification\footnote{From now on we ignore the over all constant $c_1 Y^{2 \nu}$ appearing in eq.~(\ref{eq:psi_RC}); this multiplicative constant will be restored at the end while presenting physical results.}
\begin{align}
u=(x_1 x_2)^{\alpha_1} \, (T_{1,1} T_{1,2}T_{2,1} T_{2,2})^{\alpha_2},
\label{eq:u_def_massless}
\end{align}
with
\begin{equation}
\alpha_1=\frac{1}{2}-\nu+\epsilon, \qquad \alpha_2=\nu-\frac{1}{2}. 
\end{equation}
The elements $T_{i,j}$, with $1 \leq i,j \leq2$ are hyperplanes defined by
\begin{equation}
\begin{aligned}
T_{1,1}=t_1-1,\qquad
T_{1,2}=t_1+1,\qquad
T_{2,1}=t_2-1,\qquad
T_{2,2}=t_2+1.     
\end{aligned}
\end{equation}
The differential form associated to the right-center term eq.~(\ref{eq:psi_RC}) is\footnote{Notice that in comparison to eq.~(\ref{eq:psi_RC}) we absorbed a factor of $(x_1x_2)^{-1}$ into the differential form.} 
\begin{equation}
    \varphi_{RC}=\frac{dx_1 \wedge dx_2 \wedge dt_1 \wedge dt_2}{x_1 x_2\left(k_{12}+k_{34}-t_1 Y+t_2 Y+x_1+x_2\right)\left(k_{34}+t_2 Y+x_2\right) }.
    \label{eq:phi_LC}
\end{equation}
For the later convenience, we define two additional hyperplanes $S_1$ and $S_2$, appearing in eq.~(\ref{eq:phi_LC})
\begin{equation}
\begin{aligned}
S_1=k_{12}+k_{34}-t_1 Y+t_2 Y+x_1+x_2, \quad
S_2=k_{34}+t_1 Y+x_1.  
\end{aligned}    
\label{eq:S_i_j_definition}
\end{equation}
Considering the definition of $u$ in eq.~(\ref{eq:u_def_massless}) and $\mathcal{C}=(0, \infty)^2 \times (1, \infty)^2$, we can view eq.~(\ref{eq:psi_RC}) as a member of a more general integral family defined as
\begin{equation}  \int_{\mathcal{C}} u \, \frac{dx_1 \wedge dx_2 \wedge dt_1 \wedge dt_2}{x_1^{a_1} \, x_2^{a_2} \, T_{1,1}^{a_3} \, T_{1,2}^{a_4}\, T_{2,1}^{a_5} \, T_{2,2}^{a_6} \, S_1^{a_7}\, S_{2}^{a_8}}\,, \qquad a_i \in \mathbb{Z}\, .
\label{eq:integral_family_RC}
\end{equation}

We should notice that the hyperplanes $S_1$ and $S_2$ in eq.~(\ref{eq:integral_family_RC}) are not contained in the set of hypersurfaces defining $u$, cf. eq.~(\ref{eq:u_def_massless}). In many practical examples~\cite{Frellesvig:2019uqt,Frellesvig:2020qot}, it was shown that one possible way of dealing with this situation is to consider a slight modification of the original setup. In practice, this amounts to consider the replacement
\begin{equation}
    u \to u \left( S_1 S_2 \right)^{r},
\label{eq:u_modified_regulator}
\end{equation}
in the original integral, where the parameter $r$ act as a ``regulator". One can then work with the modified version eq.~(\ref{eq:u_modified_regulator}), rather than the original eq.~(\ref{eq:u_def_massless}). It is indeed possible to derive linear relations for the modified integrals defined with eq.~(\ref{eq:u_modified_regulator}) and read them in the limit $r \to0$. In our example, a critical points analysis gives (upon the replacement eq.~(\ref{eq:u_modified_regulator}))
\begin{equation}
    \dim \operatorname{H}^4(X, \nabla_{\omega}) =8\,.
\end{equation}
Similarly, the differential forms the LC$'$ and LR terms read
\begin{equation}
\begin{aligned}
 \varphi_{LC'}&=\frac{dx_1 \wedge dx_2 \wedge dt_1 \wedge dt_2}{x_1 x_2\left(k_{12}+k_{34}+t_1 Y-t_2 Y+x_1+x_2\right)\left(k_{12}+t_1 Y+x_1\right) }\,,
\\
 \varphi_{LR}&=\frac{dx_1 \wedge dx_2 \wedge dt_1 \wedge dt_2}{x_1 x_2\left(k_{12}+t_1 Y+x_1\right) \left(k_{34}+t_2 Y+x_2\right)}\,. \label{eq:phi_RC_LR}
 \end{aligned}
\end{equation}
In this work, rather than considering the modification in eq~(\ref{eq:u_modified_regulator}), we opt for the method of relative twisted cohomology--summarized in the next Section. From a practical point of view, this approach avoids the introduction of additional regulators. 
\subsection{Short review of relative twisted cohomology} \label{sec:relative_coh}
As we mentioned in the previous Section in the mathematical setup of twisted cohomology all the hyperplanes in the integrand should appear in $u$ eq.~(\ref{eq:u_def_massless}) (colloquially we say that all the possible singularities are ``regulated" by the twist $u$). However, as we noticed above, this condition is not met by the integrals we are investigating. It turns out that the proper mathematical framework for tackling the problem, is given by (dual) relative twisted cohomology. In this space we can identify in a natural way a basis for the cohomology and derive differential equations obeyed by the latter. We can then exploit an isomorphism between the dual and the ``direct" cohomologies, to establish the differential equations for the problem of our interest, namely for the basis of the cohomology associated to eq.~(\ref{eq:integral_family_RC}).\\
In this section we provide a short and concise introduction to relative twisted cohomology from a practical stand point; we follow closely the discussion in~\cite{De:2023xue}.\\
~\\
For the class of integrals considered in this work, let us introduce
\begin{equation}
\begin{split}
    & S_i: \text{Singular hypersurface associated to } (\varphi u) \text{ with integer power}\, ,\\ 
\end{split}
\end{equation}
where $u$ is the same as in eq.~(\ref{eq:hyperint}). Assuming that we have a collection of $\ell$ singular hypersurfaces $\{S_1, \dots, S_{\ell}\}$, referred to as ``boundaries", we define
\begin{equation}
\begin{split}
    X^\vee= \mathbb{C}^n \setminus \{ (\cup_{i=1}^{m}\, l_i=0) \}\,, \qquad S^\vee= X^{\vee} \cap\left( \cup^{\ell}_{i=1} S_i=0 \right)\,.
\end{split}
\end{equation}
Moreover we introduce the dual connection
\begin{equation}
\nabla^{\vee}(\bullet)=d(\bullet)+\omega^\vee \wedge \bullet, \qquad
\text{with: $\omega^\vee=d \log(u^{-1})= - d \log (u)\,.$}
\end{equation}
Colloquially speaking, (dual) relative twisted cohomology, denoted via $\operatorname{H}^n(X^{\vee}, S^{\vee}, \nabla^{\vee})$, is the mathematical framework that describes Stokes' theorem in presence of relative boundary terms\footnote{This is in contrast to eq.~(\ref{eq:IBP_twisted_cohomolohy}).}, i.e. differential forms supported on the $S_i$ and their intersections. Within this framework we can think of a generic differential form, say $\varphi^{\vee}$, as a formal sums of the following terms
\begin{equation}
    \theta \varphi^{\vee}_{\emptyset}, \quad \delta_i \left( \theta \varphi^{\vee}_i \right), \quad \delta_{i_1 i_2} \left( \theta \varphi^{\vee}_{i_1 i_2}  \right), \quad \dots \, .
\label{eq:formal_sum_theta_delta}
\end{equation}
Some explanations are in order; each $\varphi^{\vee}_i$ is a differential form defined on $X^{\vee} \cap S_i$ (resp. $ \varphi^{\vee}_{i_1 i_2}$ is defined on $X^{\vee} \cap S_{i_1} \cap S_{i_2}$), while $\varphi^{\vee}_{\emptyset}$ is defined on $X^{\vee}$, i.e. it is not restricted on any $S_i$.

The symbol $\delta_i(\bullet)$ localizes a differential form on $X^{\vee} \cap S_i$ (resp. $\delta_{i_1 i_2}(\bullet)$ localizes on $X^{\vee} \cap S_{i_1} \cap S_{i_2}$) and inherits the anti-commuting properties of a $1-$form (resp. $2-$form). Finally we have $\theta= \prod_{i=1}^{\ell} \theta( S_i)$ where each $\theta(S_i)$ works as a step function, i.e. it is vanishing in a small tubular neighborhood of $S_i$. Its purpose consists in (automatically) generating the boundary terms upon the action of a differential operator. Through this work we often omit writing explicitly the symbol $\theta$. Formal manipulations, relevant for this work, involving the objects we introduced are recalled in Appendix \ref{app:algebra_operators}.

The key point is that, in order to study dual relative twisted cohomology, it is sufficient to study the dual twisted cohomology on each possible subspace, i.e. $X^{\vee} \cap S_i$, $X^{\vee} \cap S_{i_1} \cap S_{i_2}$ and so on.

In particular, once we identify a basis for the dual twisted cohomology on each subspace, we can obtain a basis for the dual relative twisted cohomology simply endowing differential forms with the $\delta_i(\bullet)$, $\delta_{i_1 i_2}(\bullet), \dots ,$ introduced above, and considering (formal) sums of these quantities~\cite{Caron-Huot:2021xqj}.

Let us now comment on the isomorphism between dual and direct cohomologies. Given $\delta_i(\phi^\vee)$ a basis element of the (dual) relative cohomology, we can construct a (possible) basis of the direct cohomology via the replacement
\begin{equation}
    \delta_i \to d \log S_i\,. 
\label{eq:map_to_direct}
\end{equation} 
There exists a pairing between the two cohomologies known as intersection number~\cite{matsumoto1998}, which we denote by
\begin{equation}
    \mathbf{C}_{mn}=\langle \delta(\phi_m^\vee)| \phi_n \rangle\,;
    \label{eq:C_mat_def}
\end{equation}
for explicit formulae for the actual evaluation of eq.~(\ref{eq:C_mat_def}), relevant for the cases of our interest in this work, we refer to appendix A of~\cite{De:2023xue}; other different and more exhaustive approaches can be found in \cite{Brunello:2023rpq,Brunello:2024tqf}. Given the differential equation for the dual basis denoted as $\mathbf{A}^{\vee}$ the corresponding differential equation for the direct basis reads\footnote{Notice that $\mathbf{A}$ is independent of overall sign in $\mathbf{C}.$}~\cite{Caron-Huot:2021iev,Caron-Huot:2021xqj}
\begin{equation}
    \mathbf{A}=-(\mathbf{C}^{-1} \cdot \mathbf{A}^\vee \cdot \mathbf{C} )^{T}\, .
\label{eq:todirecto_cohomology}    
\end{equation}
\subsubsection{An example of the relative twisted cohomology framework}\label{subsec:example_LR}
\begin{figure}[t]
    \centering
    \includegraphics[scale=.25]{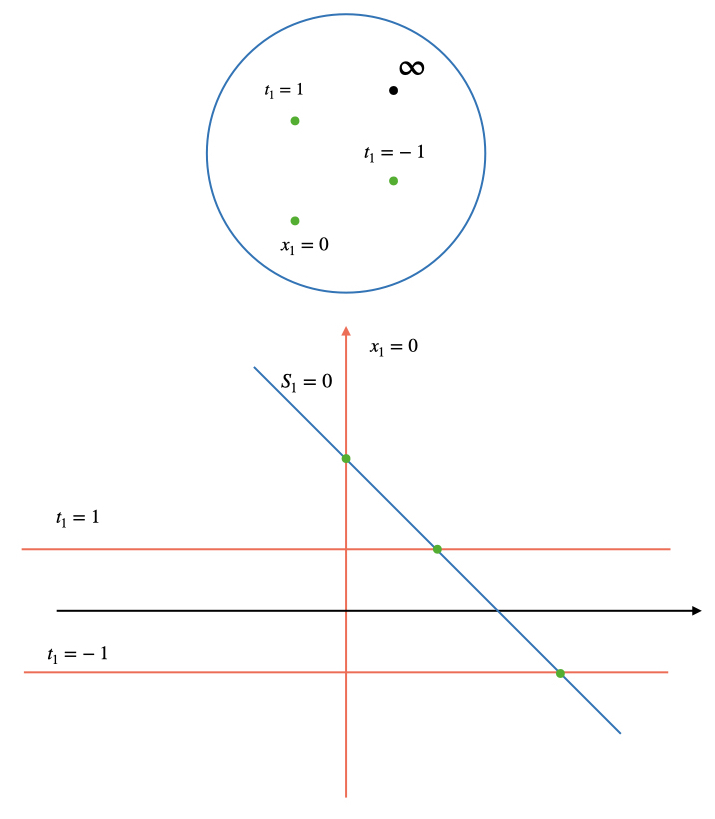}
    \caption{Hypersurfaces associated to eq.~(\ref{eq:psi_LR}). The hypersurface $S_1$ is depicted in blue.}
    \label{bg3}
\end{figure}
We consider here a pedagogical example in order to familiarise and apply the concepts introduced in the previous Section. We consider the integral in eq.~(\ref{eq:psi_LR}); this integral factorizes into a ``left part" (associated to the integration variables $x_1$ and $t_1$) discussed hereafter, and a ``right part" (associate to the integration variables $x_2$ and $t_2$). Omitting for ease of notation the over-all constants, we consider
\begin{equation}
    \psi^{L}_2(k_{12},Y)=\int_{0}^{\infty} dx_1       \int_1^{\infty}d t_1    \frac{ x_1^{\frac{1}{2}+\epsilon-\nu} (t_1^2-1)^{\nu-1/2}  }{x_1\left(k_{12}+t_1 Y+x_1\right)}\,,
\label{eq:psi_LR_new}
\end{equation}
where for later convenience we introduce
\begin{equation}
   \begin{split}
   & T_1=t_1+1\,, \quad T_2=t_1-1\,,  \qquad
    u=x_1^{\alpha_1}(T_1 T_2)^{\alpha_2}\,,
    \end{split}
\end{equation}
with 
\begin{equation}
\alpha_1=\frac{1}{2}{+} \epsilon {-} \nu\,, \qquad \alpha_2= \nu - \frac{1}{2}\,.
\end{equation}
Hence we have
\begin{equation}
X^{\vee}=\mathbb{C}^2 \setminus\{x_1=0 \,\cup \,T_1=0\, \cup\, T_2=0\}\,.
\end{equation}
In this example we have only one singular hyperplane
\begin{equation}
S_1=\{k_{12}+t_1 Y+x_1=0\}\,.
\end{equation}
The twisted cohomology on $X^{\vee}$ is zero dimensional; this is compatible with the fact that $x_1,T_1$ and $T_2$ do not form any closed chamber (cf.~Figure \ref{bg3}). 
We can see that the subspace $S^{\vee}=X^{\vee} \cap S_1$ is given by a four punctured sphere, and therefore the twisted cohomology on this space is two dimensional (cf.~Figure \ref{bg3}). We can write the two basis elements of the dual twisted cohomology $\operatorname{H}^1( S^{\vee},\nabla^{\vee}_{1})$ as\footnote{The symbol $\bullet|_1$ stands for the restriction on $X^{\vee} \cap S_1$. Similarly $\nabla^{\vee}_1 = \nabla^{\vee}|_1$.}
\begin{equation}
    \phi^\vee_{1,1}=d \log \left( \frac{T_1|_{1}}{x_1|_{1}} \right)\,,  \qquad
    \phi^\vee_{1,2}=d\log\left(\frac{T_2|_{1}}{x_1|_{1}} \right)\,.
\end{equation}
Since we have only one hyperplane we do not need to go any further and a basis for the full dual relative twisted cohomology $\operatorname{H}^2(X^\vee, S^\vee,\nabla^{\vee})$ is given by
\begin{equation}
      \delta_1(\phi_{1,1}^\vee)=\delta_1 d \log \left( \frac{T_1|_{1}}{x_1|_{1}} \right) \,, \qquad
       \delta_1 (\phi_{1,2}^\vee)=\delta_1 d\log\left( \frac{T_2|_{1}}{x_1|_{1}} \right) \,.
 \label{eq:basis_toy_model}
\end{equation}
As a simple example we can construct the differential equations associated to these basis 
differentials
\begin{equation}
    \begin{aligned}
        & \nabla^{\vee}_{\text{kin}} \delta_1(\phi_{1,i}^\vee)
        =\sum_{j=1}^2\mathbf{A}^{\vee}_{ij}\wedge \delta_1(\phi_{1,j}^\vee)\,.
    \end{aligned}
\label{eq:DEQ_toy_model}
\end{equation}
In order to derive $\mathbf{A}^{\vee}_{ij}$ we need two steps: first we have to explain how the operator $\nabla^\vee_{\text{kin}}$ acts on a generic element of the basis eq.~(\ref{eq:basis_toy_model}), and second we need to add to the l.h.s. in eq.~(\ref{eq:DEQ_toy_model}) a suitable IBP relation (i.e. $\nabla^{\vee}$ acting on a dual form $\delta_{1}(\xi^{\vee}_i)$, where $\xi_i$ has to be properly constructed) in order to read the l.h.s. in eq.~(\ref{eq:DEQ_toy_model}) as a linear combination of basis elements, and therefore reconstruct $\mathbf{A}^{\vee}_{ij}$. Taking into account the algebra of the operators $\nabla^{\vee}_{\text{kin}}$ and $\nabla^{\vee}$ as reviewed in Appendix~\ref{app:algebra_operators}, we can rewrite eq.~(\ref{eq:DEQ_toy_model}) as
\begin{equation}
\begin{split}
    \nabla^{\vee}_{\text{kin}} \delta_1 (\phi^{\vee}_{1,i}) & \cong \nabla^{\vee}_{\text{kin}} \delta_1 (\phi^{\vee}_{1,i})+\nabla^{\vee} \delta_1 (\xi_i) \\
    & =\delta_{1} \left( \nabla^{\vee}_{\text{kin}}|_{1} \phi^{\vee}_{1,i}      \right)-\delta_{1} \left( \nabla^{\vee}_{1} \xi_i \right) \\
     & =\sum_{j=1}^2\mathbf{A}^{\vee}_{ij}\wedge \delta_1(\phi_{1,j}^\vee)\,.
\label{eq:DEQ_toy_model_cong}
\end{split}
\end{equation}
In order to illustrate a concrete example, let us focus on the $k_{12}-$component of $\nabla^{\vee}_{\text{kin}}|_{1}$, dubbed as $\nabla^{\vee}_{k_{12}}|_{1}$, acting on the first basis element $\phi^{\vee}_{1,1}$; we have 
\begin{equation}
    \nabla^{\vee}_{k_{12}}|_{1} \phi^{\vee}_{1,1}
    =\frac{Y(\alpha _1
   +1) \, d t_1}{\left(k_{12}+t_1
   Y\right){}^2}-\frac{\alpha
   _1\, d t_1}{(t_1-1)
   \left(k_{12}+t_1
   Y\right)}\,;
\label{eq:k12_componenet_nabla_kin}
\end{equation}
we notice that the first term in eq.~(\ref{eq:k12_componenet_nabla_kin}) has a double pole. Our goal is to construct a suitable $0-$form $\xi_1$ in order to remove this term. Let us consider the ansatz
\begin{equation}
    \xi_1= \frac{c_1}{\left(k_{12}+t_1
   Y\right)}\,,
\end{equation}
then
\begin{equation}
\begin{split}
\nabla^{\vee}_{k_{12}}|_1 \phi^{\vee}_{1,1}-\nabla^{\vee}_1 \xi_1 & = \frac{\left(\alpha _1+1\right) \left(c_1+1\right)
   Y\, d t_1}{\left(k_{12}+t_1 Y\right){}^2}+\frac{\alpha
   _2 c_1 \, d t_1}{\left(t_1+1\right) \left(k_{12}+t_1
   Y\right)}\\
   & +\frac{(\alpha _2 c_1-\alpha
   _1) \, dt_1}{\left(t_1-1\right) \left(k_{12}+t_1
   Y\right)}\,;
\end{split}
\label{eq:double_poles_expression}
\end{equation}
we infer that eq.~(\ref{eq:double_poles_expression}) is free from double poles provided the choice $c_1=-1$. We can then read eq.~(\ref{eq:double_poles_expression}) (for $c_1=-1$) as a linear combination of basis elements, therefore obtaining the first row of the $k_{12}-$component of $\mathbf{A}^{\vee}_{ij}$. The explicit elements read
\begin{equation}
    \mathbf{A}^{\vee}_{k_{12},11} ={-}\frac{\alpha _1{+}\alpha _2}{k_{12}{+}Y}\,, \qquad \mathbf{A}^{\vee}_{k_{12},12}={-}\frac{\alpha _2}{k_{12}{-}Y}\,.
\end{equation}
Proceeding in a similar way we can obtain the full $\mathbf{A}^{\vee}$, whose explicit expression reads
\begin{equation}
 \mathbf{A}^{\vee}=\begin{pmatrix}
        & \alpha_2r_1-(\alpha_1{+}\alpha_2)r_2 & \alpha_2(r_1- r_3) \\
        & \alpha_2 (r_1- r_2) & \alpha_2 r_1-(\alpha_1{+}\alpha_2)r_3
    \end{pmatrix}\,,
\end{equation}
where $r_i$ are defines as 
\begin{equation}
    \begin{aligned}
        r_1&=d \log Y\,,\\
        r_2&=d \log (k_{12}+Y)\,,\\
        r_3&=d \log (k_{12}-Y)\,.\\
    \end{aligned}
\end{equation}
Given these dual basis elements we can now construct the basis for the direct cohomology  $\operatorname{H}^2(X,\nabla)$ as
\begin{equation}
    \begin{aligned}
     & \phi_1=d \log \left( {T_1} \right) \wedge d \log S_1\,,\\
      & \phi_{2}=d \log \left( {T_2} \right) \wedge d \log S_1\,.
 \end{aligned}
\end{equation}
As we discussed in the previous Section we need the explicit expression for the matrix $\mathbf{C}$ in order to map the matrix $\mathbf{A}^{\vee}$ to the direct cohomology matrix $\mathbf{A}$. For the current example we have the intersection matrix as 
\begin{equation}
    \mathbf{C}_{ab}=\frac{1}{\alpha_2} \delta_{ab}\,,
\end{equation}
and therefore the matrix $\mathbf{A}$ (in this particular case) reads 
\begin{equation}
    \mathbf{A}=-(\mathbf{A}^\vee)^T\,.
\end{equation}
\subsection{Relative twisted cohomology of internally massive correlator with conformally coupled external states}
\begin{figure}[h]
    \centering
    \includegraphics[scale=.30]{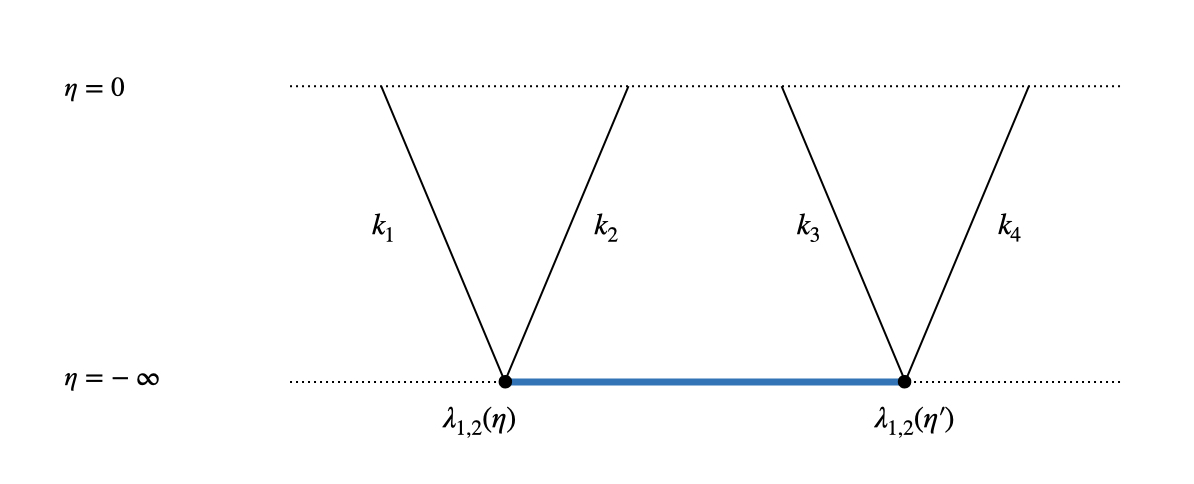}
    \caption{Two--site tree level correlator with one massive internal state}
    \label{diag}
\end{figure}
After formal preparations we are now ready to discuss the two--site correlator with massive internal propagator. As we mentioned at the end of Section \ref{sec2}, our goal is to compute the integral defined in eq.~(\ref{eq:psi_RC}) (other contributions can be obtained either by symmetry or by direct integration). The dual relative twisted cohomology  can be defined via the twist
\begin{equation}
    \begin{aligned}
    & u=(x_1x_2)^{\alpha_1 }(T_{1,1}T_{1,2} T_{2,1}T_{2,2})^{\alpha_2 }=(x_1x_2)^{\alpha_1 }((t_1-1)(t_1+1)(t_2-1)(t_2+1))^{\alpha_2}\,.\\
     \end{aligned}
\end{equation}
Hence we have the following space
\begin{equation}
    \begin{aligned}
         & X^\vee=\mathbb{C}^4\setminus\{x_1=0 \cup x_2=0 \cup T_{1,1}=0\cup T_{1,2}=0\cup T_{2,1}=0 \cup T_{2,2}=0\}\,.\\
    \end{aligned}
\end{equation}
The singular hypersurfaces are given by (the vanishing of)
\begin{equation}
S_1=k_{12}+k_{34}-t_1 Y+t_2 Y+x_1+x_2\,, \qquad
S_2=k_{34}+t_2 Y+x_2\,.
\end{equation}
The first step is to compute the dimension and the basis elements of the dual relative twisted cohomology. The cohomology on $X^{\vee}$ is zero dimensional (as was the case for the previous example). Then, we consider the cohomology restricted to the subspace $X^{\vee} \cap S_2$; we have the following system defined by the restricted twist $u|_{2}$ (we choose to remove the variable $x_2$):
\begin{equation}
   \begin{aligned}
     u|_{2}&=(x_1)^{\alpha_1 }(-k_{34}-t_2 Y)^{\alpha_1 }(T_{1,1}T_{1,2} T_{2,1}T_{2,2})^{\alpha_2 },\\
    \omega|_{2}& =d\log u|_{2}.\\  
   \end{aligned}
\end{equation}
The number of critical points of this subspace is zero and hence the cohomology has zero dimension and it is trivial. This fact can also be geometrically predicted since the removed hyperplanes do not produce any closed chambers. 

The second subcohomology corresponding to $X^{\vee} \cap S_1$ is associated to the following system given by the restricted twist $u|_1$ (we choose to remove variable $x_1$)
\begin{equation}
   \begin{aligned}
    u|_{1}&=(-k_{12}-k_{34}+t_1 Y-t_2 Y-x_2 )^{\alpha_1 }(x_2)^{\alpha_1 }(T_{1,1}T_{1,2} T_{2,1}T_{2,2})^{\alpha_2 },\\
    \omega|_{1}& =d\log u|_{1};\\
   \end{aligned}
\end{equation}
a critical point analysis gives a four dimensional cohomology on $X^\vee \cap S_1$.

Finally, we have the third subspace associated to the intersection of these two subspaces i.e. $X^{\vee} \cap S_1 \cap S_2$; a critical point analysis indicate a four dimensional cohomology. Together with the subcohomology on $X^\vee \cap S_1$ we obtain, in total, an eight dimensional dual relative twisted cohomology for the full system.
Choosing the variables $\{x_2,t_1,t_2\}$ to parameterize $X^{\vee} \cap S_1$, then the basis elements $\{\phi^\vee_i\}_{i=1}^4$ on this subspace can be chosen as
\begin{equation}
{\footnotesize
\begin{aligned}
  & \phi^\vee_1= d \log \left(\frac{T_{1,1}|_1}{T_{2,1}|_1} \right) \wedge d \log \left(\frac{T_{2,1}|_1}{x_1|_1} \right)\wedge  d\log \left( \frac{x_1|_1}                                             {x_2|_1} \right)=-\frac{(k_{12}{+}k_{34})\,\,dt_1 \wedge dt_2 \wedge dx_{2}}{(t_1{-}1)(t_2{-}1)(k_{12}{+}k_{34}{-}t_1 Y{+}t_2 Y{+}x_2)x_2},       \\
                                   & \phi^\vee_2= d \log \left(\frac{T_{1,1}|_1}{T_{2,2}|_1}\right) \wedge d \log\left(\frac{T_{2,2}|_1}{x_1|_1} \right)\wedge  d\log \left( \frac{x_1|_1}{x_2|_1} \right)=-\frac{(k_{12}+k_{34}{-}2Y)\,\,dt_1 \wedge dt_2 \wedge dx_{2}}{(t_1{-}1)(t_2{+}1)(k_{12}{+}k_{34}{-}t_1 Y{+}t_2 Y{+}x_2)x_2},       \\
                                   & \phi^\vee_3= d \log \left( \frac{T_{1,2}|_1}{T_{2,1}|_1} \right) \wedge d \log \left( \frac{T_{2,1}|_1}{x_1|_1} \right)\wedge  d\log \left(\frac{x_1|_1}{x_2|_1} \right)=-\frac{(k_{12}{+}k_{34}{+}2Y)\,\,dt_1 \wedge dt_2 \wedge dx_{2}}{(t_1{+}1)(t_2{-}1)(k_{12}{+}k_{34}{-}t_1 Y{+}t_2 Y{+}x_2)x_2},      \\
                                   & \phi^\vee_4= d \log \left( \frac{T_{1,2}|_1}{T_{2,2}|_1} \right) \wedge d \log \left( \frac{T_{2,2}|_1}{x_1|_1} \right)\wedge  d\log \left( \frac{x_1|_1}{x_2|_1} \right) =-\frac{(k_{12}{+}k_{34})\,\,dt_1 \wedge dt_2 \wedge dx_{2}}{(t_1{+}1)(t_2{+}1)(k_{12}{+}k_{34}{-}t_1 Y{+}t_2 Y{+}x_2)x_2}\,.   \\
                             \\     \label{basisa}        
                             \end{aligned}}
\end{equation}
Similarly choosing $\{t_1,t_2 \}$ to parameterize the subspace $X^{\vee} \cap S_1 \cap S_2$, the differential forms $\{\phi^{\vee}_i\}_{i=5}^8$ associated to it can be chosen as 
\begin{equation}
{\footnotesize
\begin{aligned}
      & \phi^\vee_5= d \log \left(\frac{x_1|_{12}}{T_{1,1}|_{12}} \right) \wedge d \log \left( \frac{x_2|_{12}}{T_{2,1}|_{12}} \right)=-\frac{(k_{12}-Y)(k_{34}+Y)dt_1 \wedge dt_2}{(t_1-1)(t_2-1)(t_1 Y-k_{12})(t_2 Y+k_{34})},       \\
                                   & \phi^\vee_6= d \log \left( \frac{x_1|_{12}}{T_{1,1}|_{12}} \right) \wedge d \log \left( \frac{x_2|_{12}}{T_{2,2}|_{12}} \right)=-\frac{(k_{12}-Y)(k_{34}-Y)dt_1 \wedge dt_2}{(t_1-1)(t_2+1)(t_1 Y-k_{12})(t_2 Y+k_{34})},     \\
                                   & \phi^\vee_7= d \log \left( \frac{x_1|_{12}}{T_{1,2}|_{12}} \right) \wedge d \log \left( \frac{x_2|_{12}}{T_{2,1}|_{12}} \right)=-\frac{(k_{12}+Y)(k_{34}+Y)dt_1 \wedge dt_2}{(t_1+1)(t_2-1)(t_1 Y-k_{12})(t_2 Y+k_{34})},         \\
                                   & \phi^\vee_8= d \log \left( \frac{x_1|_{12}}{T_{1,2}|_{12}} \right) \wedge d \log \left( \frac{x_2|_{12}}{T_{2,2}|_{12}} \right)=\frac{(k_{12}+Y)(Y-k_{34})dt_1 \wedge dt_2}{(t_1+1)(t_2+1)(t_1 Y-k_{12})(t_2 Y+k_{34})}.     \\
                              \\   \label{basisb}  
                              \end{aligned}}
\end{equation}
Combining these two sets together we have the eight basis elements of the full dual relative twisted cohomology as 
\begin{equation}
\begin{split}
     \{\delta_{1}(\phi^\vee_1),\delta_{1}(\phi^\vee_2),\delta_{1}(\phi^\vee_3),\delta_{1}(\phi^\vee_4),\delta_{12}(\phi^\vee_5),\delta_{12}(\phi^\vee_6),\delta_{12}(\phi^\vee_7),\delta_{12}(\phi^\vee_8)\}\,.
\label{eq:dual_basis}
\end{split}
   \end{equation}
Our goal is to find the matrix $\mathbf{A}^{\vee}_{ij}$ fulfilling
\begin{equation}
    \begin{aligned}
        & \nabla^\vee_{\text{kin}}  \delta_{\bullet} (\phi^\vee_i)\cong \nabla^\vee_{\text{kin}}  \delta_{\bullet} (\phi^\vee_i) + \nabla^\vee  \delta_{\bullet} (\xi^\vee_i) = \sum_{j=1}^8 \mathbf{A}_{ij}^{\vee} \wedge    \delta_{\bullet}( \phi^\vee_j)\,,\\
         \end{aligned} \label{Aeq}
\end{equation}
where $\delta_{\bullet}=\delta_1$ if it acts on $\phi^{\vee}_i$ with $1 \leq i \leq 4$ and $\delta_{\bullet}=\delta_{12}$ otherwise.\\

~\\
Let us consider first a generic basis element of the form $\delta_{12}(\phi^\vee_i)$, with $5 \leq i \leq 8$ (those are simpler to understand). Let $\xi_i^{\vee}$ be a  $1-$form w.r.t. the kinematic variables and a $1-$form w.r.t. the integration variables defined on $X^{\vee} \cap S_1 \cap S_2$.\\
In this case eq.~(\ref{Aeq}) reads (using $\nabla^{\vee} \delta_{12}(\xi^{\vee}_i)=\delta_{12} \left( \nabla^{\vee}|_{12}\, \xi^{\vee}_i    \right)$)  
\begin{equation}
\begin{aligned}
        &  \delta_{12}\Bigg( \nabla^{\vee}_{\text{kin
}}|_{12}\,\phi^\vee_i+\nabla^\vee|_{12} \, \xi^\vee_i-\sum_{j=5}^8 \mathbf{A}_{ij}^{\vee} \wedge   \phi^\vee_j\Bigg)=0\,, 
\qquad \qquad 5 \leq i \leq 8\,.
    \end{aligned}
    \label{Aeq2}
\end{equation}
Let us now come to a basis element of the form $\delta_1(\phi^{\vee}_i)$ with $1 \leq i \leq 4$. In this case, we denote with $\xi_i^{\vee}$ a $1-$form w.r.t. the kinematics and a $2-$form w.r.t. the integration variables defined on $X^{\vee} \cap S_1$ (using $\nabla^{\vee} \delta_1( \xi^{\vee}_i)=-\delta_1 (\nabla^{\vee}|_{1} \xi^{\vee}_i) - \delta_{12} (\xi^{\vee}_i|_{2}$))
\begin{equation}
    \begin{aligned}
        & \delta_1\Bigg( (\nabla^\vee_{\text{kin}}|_{1}    \,\phi^\vee_i)-\nabla^\vee \xi^\vee_i-\sum_{j=1}^4 \mathbf{A}_{ij}^{\vee} \wedge   \phi^\vee_j\Bigg)-\delta_{12}\Bigg(  \xi^\vee_i|_2- d _{\text{kin}} \theta_2 \wedge  \phi^\vee_i|_{2} \\
        &+\sum_{j=5}^8 \mathbf{A}_{ij}^{\vee} \wedge   \phi^\vee_j\Bigg)=0\,, \qquad 1 \leq i \leq 4\,.\\
          \end{aligned} \label{Aeq1}
\end{equation}
The differential forms $\xi^{\vee}$ in eqs.~(\ref{Aeq2}) and~(\ref{Aeq1}) can be systematically constructed requiring that the double poles arising from the action of $\nabla^{\vee}_{\text{kin}}$ are absent, similarly to the discussion in Subsection \ref{subsec:example_LR}. As an example we report here the  $k_{12}$ component of the matrix $\mathbf{A}^{\vee}_{ij}$
 \begin{equation}\scalemath{0.9}{ 
\mathbf{A}_{k_{12}}^{\vee}=  \begin{pmatrix}
 -\frac{2 \alpha}{k_{12}+k_{34}} & -\frac{\alpha_2}{k_{12}+k_{34}-2 Y} & -\frac{\alpha_2}{k_{12}+k_{34}+2 Y} & 0 & -\frac{1}{k_{12}-Y} & 0
   & 0 & 0 \\
 -\frac{\alpha_2}{k_{12}+k_{34}} & -\frac{2 \alpha}{k_{12}+k_{34}-2 Y} & 0 & -\frac{\alpha_2}{k_{12}+k_{34}} & 0 & -\frac{1}{k_{12}-Y} & 0
   & 0 \\
 -\frac{\alpha_2}{k_{12}+k_{34}} & 0 & -\frac{2 \alpha}{k_{12}+k_{34}+2 Y} & -\frac{\alpha_2}{k_{12}+k_{34}} & 0 & 0 & -\frac{1}{k_{12}+Y}
   & 0 \\
 0 & -\frac{\alpha_2}{k_{12}+k_{34}-2 Y} & -\frac{\alpha_2}{k_{12}+k_{34}+2 Y} & -\frac{2 \alpha}{k_{12}+k_{34}} & 0 & 0 & 0 &
   -\frac{1}{k_{12}+Y} \\
 0 & 0 & 0 & 0 & \frac{\alpha}{Y-k_{12}} & 0 & -\frac{\alpha_2}{k_{12}+Y} & 0 \\
 0 & 0 & 0 & 0 & 0 & \frac{\alpha}{Y-k_{12}} & 0 & -\frac{\alpha_2}{k_{12}+Y} \\
 0 & 0 & 0 & 0 & \frac{\alpha_2}{Y-k_{12}} & 0 & -\frac{\alpha}{k_{12}+Y} & 0 \\
 0 & 0 & 0 & 0 & 0 & \frac{\alpha_2}{Y-k_{12}} & 0 & -\frac{\alpha}{k_{12}+Y} \\
 \end{pmatrix}}\,,
\end{equation}
with $\alpha=\alpha_1+\alpha_2$.

In order to obtain differential equations in direct cohomolohy we need the intersection matrix $\mathbf{C}$
\begin{equation}
    \mathbf{C}_{ab}=\langle \delta(\phi^\vee_a)|\phi_b \rangle\,,
\end{equation}
where $\phi^\vee_a$ are the set of basis in the dual-cohomology given in (\ref{basisa}) and (\ref{basisb}) and the $\phi_b$ are the basis of the direct cohomology that can be constructed orthogonal to $\phi^\vee_a$s (c.f. eq~(\ref{eq:map_to_direct})). In particular with the choice
\begin{equation}
    \begin{aligned}
          & \phi_1= d\log(x_1)\wedge d \log(T_{1,1}) \wedge d \log(T_{2,1})\wedge   d \log(S_1)\,,    \\
          & \phi_2=d\log(x_1)\wedge d \log(T_{1,1}) \wedge d \log(T_{2,2})\wedge   d \log(S_1)\,,   \\
          & \phi_3= d\log(x_1)\wedge d \log(T_{1,2}) \wedge d \log(T_{2,1})\wedge   d \log(S_1)\,,     \\
          & \phi_4= d\log(x_1)\wedge d \log(T_{1,2}) \wedge d \log(T_{2,2})\wedge  d \log(S_1)\,,  \\   
          & \phi_5= d \log(T_{1,1}) \wedge d \log(T_{2,1})\wedge d \log(S_2) \wedge d \log(S_1)\,,       \\
          & \phi_6=d \log(T_{1,1}) \wedge d \log(T_{2,2})\wedge d \log(S_2) \wedge d \log(S_1)\,,     \\
          & \phi_7= d \log(T_{1,2}) \wedge d \log(T_{2,1})\wedge d \log(S_2) \wedge d \log(S_1),      \\
          & \phi_8=d \log(T_{1,2}) \wedge d \log(T_{2,2})\wedge d \log(S_2) \wedge d \log(S_1)\,,     \\                      
    \end{aligned}
\label{eq:basis_massless}
\end{equation}
the intersection matrix $\mathbf{C}_{ab}$ computes to: 
\begin{equation}
  \mathbf{C}_{ab} =\begin{cases}
     &-\frac{1}{\alpha_1\alpha_2^2}\delta_{ab} \qquad \qquad 1\leq a \leq 4\,,\\
     &\,\,\,\,\frac{1}{\alpha_2^2}\delta_{ab}         \qquad \qquad \,\,\,\, 5\leq a \leq 8\,.
  \end{cases}
\end{equation}
Therefore we have the following matrix in the direct cohomology (see eq.~(\ref{eq:todirecto_cohomology}))
 \begin{equation}\scalemath{0.9}{ 
\mathbf{A}_{k_{12}}=\begin{pmatrix}
 \frac{2 \alpha}{k_{12}+k_{34}} & \frac{\alpha_2}{k_{12}+k_{34}} & \frac{\alpha_2}{k_{12}+k_{34}} & 0 & 0 & 0 & 0 & 0 \\
 \frac{\alpha_2}{k_{12}+k_{34}-2 Y} & \frac{2 \alpha}{k_{12}+k_{34}-2 Y} & 0 & \frac{\alpha_2}{k_{12}+k_{34}-2 Y} & 0 & 0 & 0 & 0 \\
 \frac{\alpha_2}{k_{12}+k_{34}+2 Y} & 0 & \frac{2 \alpha}{k_{12}+k_{34}+2 Y} & \frac{\alpha_2}{k_{12}+k_{34}+2 Y} & 0 & 0 & 0 & 0 \\
 0 & \frac{\alpha_2}{k_{12}+k_{34}} & \frac{\alpha_2}{k_{12}+k_{34}} & \frac{2 \alpha}{k_{12}+k_{34}} & 0 & 0 & 0 & 0 \\
 -\frac{\alpha_1}{k_{12}-Y} & 0 & 0 & 0 & \frac{\alpha_1+\alpha_2}{k_{12}-Y} & 0 & \frac{\alpha_2}{k_{12}-Y} & 0 \\
 0 & -\frac{\alpha_1}{k_{12}-Y} & 0 & 0 & 0 & \frac{\alpha_1+\alpha_2}{k_{12}-Y} & 0 & \frac{\alpha_2}{k_{12}-Y} \\
 0 & 0 & -\frac{\alpha_1}{k_{12}+Y} & 0 & \frac{\alpha_2}{k_{12}+Y} & 0 & \frac{\alpha_1+\alpha_2}{k_{12}+Y} & 0 \\
 0 & 0 & 0 & -\frac{\alpha_1}{k_{12}+Y} & 0 & \frac{\alpha_2}{k_{12}+Y} & 0 & \frac{\alpha_1+\alpha_2}{k_{12}+Y} \\
\end{pmatrix}}\,. 
\end{equation}
Performing the calculation for all the kinematic variables we obtain the differential equation eq.~(\ref{eq:DEQ_generic}); the explicit expression of the matrix $\mathbf{A}$ is 
{\footnotesize
\begin{equation}\scalemath{0.7}{ 
\mathbf{A}=  \begin{pmatrix}
  2 r_2 \alpha-2 \alpha_2 r_1 & \alpha_2 (r_2-r_1) & \alpha_2 (r_2-r_1) & 0 & 0 & 0 & 0 & 0 \\
 \alpha_2 (r_8-r_1) & 2 r_8 \alpha-2 \alpha_2 r_1 & 0 & \alpha_2 (r_8-r_1) & 0 & 0 & 0 & 0 \\
 \alpha_2 (r_7-r_1) & 0 & 2 r_7 \alpha-2 \alpha_2 r_1 & \alpha_2 (r_7-r_1) & 0 & 0 & 0 & 0 \\
 0 & \alpha_2 (r_2-r_1) & \alpha_2 (r_2-r_1) & 2 r_2 \alpha-2 \alpha_2 r_1 & 0 & 0 & 0 & 0 \\
 \alpha_1(r_5-r_4) & 0 & 0 & 0 & \alpha (r_4+r_5)-2 \alpha_2 r_1 & \alpha_2 (r_5-r_1) & \alpha_2
   (r_4-r_1) & 0 \\
 0 &\alpha_1(r_6-r_4) & 0 & 0 & \alpha_2 (r_6-r_1) & \alpha (r_4+r_6)-2 \alpha_2 r_1 & 0 & \alpha_2
   (r_4-r_1) \\
 0 & 0 & \alpha_1(r_5-r_3) & 0 & \alpha_2 (r_3-r_1) & 0 & \alpha (r_3+r_5)-2 \alpha_2 r_1 & \alpha_2
   (r_5-r_1) \\
 0 & 0 & 0 & \alpha_1(r_6-r_3) & 0 & \alpha_2 (r_3-r_1) & \alpha_2 (r_6-r_1) & \alpha (r_3+r_6)-2
   \alpha_2 r_1 \\
\end{pmatrix}}\,, \label{eq:def_eq_A}
\end{equation}
}
where $\alpha=\alpha_1+\alpha_2$, and $r_i$ are defined as
\begin{equation}
\begin{aligned}
r_1&= d \log Y \, ,\qquad \qquad  \qquad 
 \quad r_2=d \log(k_{12}+k_{34}) \, ,
\\
r_3&=d \log (Y+k_{12}) \, , \qquad \qquad r_4=d \log(Y-k_{12}) \, ,
\\
r_5&=d \log (Y+k_{34}) \, , \qquad \qquad r_6=d \log(Y-k_{34}) \, ,
\\
r_7&=d \log (2Y+k_{12}+k_{34}) \,, \quad
r_8=d \log(2Y-k_{12}-k_{34}) \, .
\\ 
\label{eq:letter}
\end{aligned}
\end{equation}
In the next Section we will verify that these set of differential equations can also be produced through direct IBP considerations. 

 Finally, in order to obtain the final result for the correlator eq.~(\ref{eq:correlator_complete}), we should express the integral associated to eq.~(\ref{eq:phi_LC}) in terms of MIs. The explicit decomposition reads
\begin{equation}
    I(\varphi_{RC})=\frac{\alpha^2_2}{ \alpha^2_1 \,Y^2}\Big(\text{I}(\phi_5)+\text{I}(\phi_7)+\text{I}(\phi_7)+\text{I}(\phi_8)\Big)\,. \label{eq:deco_RC}
\end{equation}
The result for $I(\varphi_{LC'})$ can be obtained from eq.~(\ref{eq:deco_RC}) upon the replacement $k_{34} \leftrightharpoons k_{12}$ (and, as discussed above, the last term in eq.~(\ref{eq:correlator_complete}) can be obtained by direct integration).
\subsection{Differential Equations from Integration-by-parts reductions}
\label{subsec:DEQfromIBPs}
We discuss now an alternative method to derive the system of differential equations obeyed by the MIs, without resorting on the relative twisted cohomology framework. The setup is the one defined in eq.~(\ref{eq:hyperint}), and we consider the $l_i$ to be hyperplanes independent from the kinematics (cf. eq.~(\ref{eq:u_def_massless})). We further assume that a basis of MIs (equivalently a basis for the underlying twisted cohomology) is known; we assume this basis to be of the form of eq.~(\ref{eq:basis_massless}), i.e. to have logarithmic singularities along the hyperplanes $\{l_i\}$, and a subset of the hyperplanes, say $\{S_i\}$. Let us write a given basis element as
\begin{equation}
    \phi= d \log(l_1) \wedge \cdots \wedge d \log (l_{n-|S|}) \wedge d \log (S_1) \wedge \cdots \wedge d \log (S_{|S|})\,.
\label{eq:dummy_phi}
\end{equation}
The action of $\nabla_{\text{kin}}$ on eq.~(\ref{eq:dummy_phi}) reads 
\begin{equation}
\begin{aligned}
 \nabla_{\text{kin}} \phi =d_{\text{kin}} \phi=&\sum_{i=1}^{|S|}  d\log l_1 \wedge \cdots \wedge d \log (l_{n-|S|}) 
 \wedge d\log S_{1} \wedge \cdots \\ 
& \wedge d_{\text{kin}} d\log S_{i} \wedge \cdots \wedge  d\log S_{|S|}\,.
\end{aligned}
\end{equation}
We can see that $d_{\text{kin}} d\log S_{i}$ generates a double pole along the hyperplane $S_i$. This double pole can be removed by a suitable choice of an $(n-1)$-form dubbed as $\xi_{\phi}$, whose expression reads
\begin{equation}
\begin{aligned}
    \xi_{\phi}=& \sum_{i=1}^{|S|} (-1)^{n-|S|+i-1} \, d_{\text{kin}}\log S_{i} \cdot  d \log(l_1) \wedge \cdots \wedge d \log (l_{n-|S|}) \wedge d \log (S_1) \wedge \cdots \\
    & \wedge \widehat{d\log (S_i)}\wedge \cdots \wedge d \log (S_{|S|})\,. 
 \label{eq:xi_choice}
\end{aligned}
\end{equation}
The action of the covariant derivative on $\xi_{\phi}$ is
\begin{equation}
\begin{aligned}
\nabla \xi_{\phi}  = & d_{\text{kin}} \phi + \omega \wedge \xi_{\phi}\,.
\end{aligned}
\end{equation}
Given the fact that $\nabla \xi_{\phi}$ integrates to zero, we can systematically replace $d_{\text{kin}} \phi$ (i.e. $\nabla_{\text{kin}} \phi$) with $\omega \wedge \xi_{\phi}$.
The latter has no double pole along any hyperplane and can be easily mapped to a combination of the chosen basis elements.
\subsection{Differential Equations for fully massive correlator} 
\label{fullM}
\begin{figure}[h]
    \centering
    \includegraphics[scale=.30]{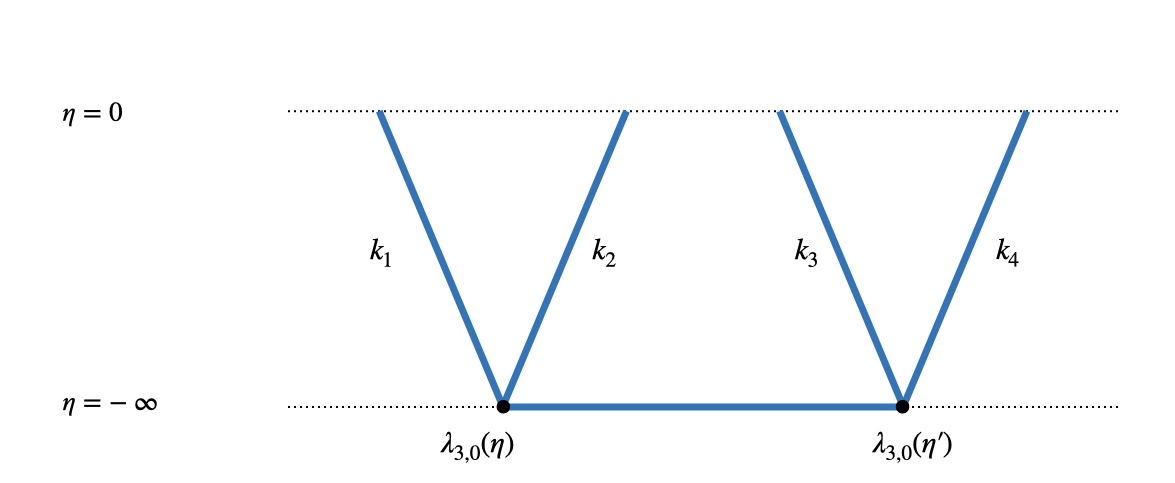}
    \caption{Two--site tree-level correlator with all massive states}
    \label{diag}
\end{figure}
In this Subsection we present the setup for the integral families related to the tree-level two site correlator with both internal and external massive states.\\
As mentioned below eq.~(\ref{eq:psi_2_massive}), the correlator is the sum of three contributions dubbed as $\psi^{LC'}_{2, \text{mass}}$, $\psi^{RC}_{2, \text{mass}}$ and $\psi^{LR}_{2,\text{mass}}$. Let us focus on $\psi^{RC}_{2, \text{mass}}$ (associated to the first line in eq.~(\ref{eq:psi_2_massive})), whose explicit expression reads
\begin{equation}
\begin{split}
    \psi^{RC}_{2,\text{mass}} = & c_2 Y^{2 \nu} \prod_{j=1}^4 k_j ^{1/2+\nu} \prod_{i=1}^2 \int_{0}^{\infty} dx_i x_i ^{-3 \nu +\epsilon -\frac{3}{2}}   \prod_{l=1}^6 \int_1^{\infty}d t_l   (t_l^2-1)^{\nu-1/2}  \times
\\
&  \frac{1}{\left(k_3 t_3{+}k_4 t_4{+}t_6 Y{+}x_2\right) \left(k_1 t_1{+}k_2 t_2{+}k_3 t_3{+}k_4 t_4{-}t_5 Y{+}t_6 Y{+}x_1{+}x_2\right)}\,.
\end{split}
\label{eq:psiRC_massive}
\end{equation}
Notice that we have the following relation:
\begin{equation} \label{eq:symmetry_mass}
\psi_{2,\text{mass}}^{LC'}(k_1,k_2,k_3,k_4,Y) = \psi_{2,\text{mass}}^{RC}(k_1,k_2,k_3,k_4,Y) \big|_{k_{i}\leftrightarrows k_{i+2}} \,\,, \qquad  i=1,2.\,
\end{equation}
We identify the twist as

\begin{equation}
          u=(x_1x_2)^{\alpha_1}(T_{1,1}T_{1,2} T_{2,1}T_{2,2}T_{3,1}T_{3,2} T_{4,1}T_{4,2}T_{5,1}T_{5,2} T_{6,1}T_{6,2})^{\alpha_2 }\,,\label{eq:Massive_twist}
\end{equation}
where
\begin{equation}
    T_{i,1}= t_i-1\,, \qquad T_{i,2}=t_i+1\,,
\label{eq:Ti_massive}
\end{equation}
and $\alpha_1$ and $\alpha_2$ read\footnote{Notice that while passing from eq.~(\ref{eq:psiRC_massive}) to eq.~(\ref{eq:Massive_twist}) we have absorbed a factor of $(x_1x_2)^{-3}$ in the differential forms. As a consequence of this, the ``physical integral" associated to eq.~(\ref{eq:psiRC_massive}) is \emph{not} an element of our basis. Its decomposition in terms of our basis elements can be achieved, for example, by means of conventional IBP reductions.}
\begin{equation}
    \alpha_1=\frac{3}{2}{+} \epsilon {-} 3\nu\,, \qquad \alpha_2= \nu - \frac{1}{2}\,.
\label{eq:alphai_massive}
\end{equation}
The integral family of our interest\footnote{Comparing to eq.~(\ref{eq:psiRC_massive}) we ignore the overall constant $c_2 Y^{2\nu} \prod\limits^4_{i=1} k_i^{\nu+1/2}$.}  is therefore (the integration domain, not relevant at this stage, is $\mathcal{C}=(0, \infty)^2 \times (1, \infty)^6$)
\begin{equation}
    \int_{\mathcal{C}} u \frac{dx_1 \wedge dx_2 \wedge dt_1 \wedge dt_2 \wedge dt_3 \wedge dt_4 \wedge dt_5 \wedge  dt_6}{x^{a_1}_1x^{a_2}_2\,T^{a_3}_{1,1}T^{a_4}_{1,2} T^{a_5}_{2,1}T^{a_6}_{2,2}T^{a_7}_{3,1}T^{a_8}_{3,2} T^{a_9}_{4,1}T^{a_{10}}_{4,2}T^{a_{11}}_{5,1}T^{a_{12}}_{5,2} T^{a_{13}}_{6,1}T^{a_{14}}_{6,2}\, S_1^{a_{15}} S_2^{a_{16}}}\,, \quad a_i \in \mathbb{Z}\,, 
\end{equation}
with
\begin{equation}
   S_1=k_3 t_3{+}k_4 t_4{+}t_6 Y{+}x_2\,,\quad S_2=k_1 t_1{+}k_2 t_2{+}k_3 t_3{+}k_4 t_4{-}t_5 Y{+}t_6 Y{+}x_1{+}x_2\,.
\label{eq:Si_massive}
\end{equation}
The corresponding dual space is
\begin{equation}
    X^\vee=\mathbb{C}^8 \setminus \{x_1=0,x_2=0,t_1\pm 1=0,t_2\pm 1=0,t_3 \pm 1=0,t_4 \pm 1=0,t_5 \pm 1=0, t_6 \pm 1 =0\}\,.
\end{equation}
We have two non zero dual relative cohomologies, one restricted on $X^{\vee} \cap S_2$ and one restricted on $X^{\vee} \cap S_1 \cap S_2$. Critical point analysis shows that each of these cohomologies is $64$ dimensional and therefore the dimension of the full twisted cohomology for the $\psi^{RC}_{2, \text{mass}}$ is $128$.\\

Concerning the ${\psi}^{LR}_{2, \text{mass}}$ contribution (associated to the last lin in eq.~(\ref{eq:psi_2_massive}))
\begin{equation}
\begin{aligned}
\psi_{2,\text{mass}}^{LR} = & c_2 Y^{2 \nu} \prod_{j=1}^4 k_j ^{1/2+\nu} \prod_{i=1}^2 \int_{0}^{\infty} dx_i x_i ^{-3 \nu +\epsilon -\frac{3}{2}}   \prod_{l=1}^6 \int_1^{\infty}d t_l   (t_l^2-1)^{\nu-1/2} \times \\
& \frac{h_0  \exp(2 i \pi \nu)}{\left(k_1 t_1{+}k_2 t_2{+}t_5 Y{+}x_1\right) \left(k_3 t_3{+}k_4 t_4{+}t_6 Y{+}x_2\right)} \,,
\end{aligned} 
\end{equation}
the integral factorizes into a left and right part and the two are once again related analogously to eq.~(\ref{eq:symmetry_mass}). We chose to derive the differential equations for the right part.\\
The twist is now 
\begin{equation}
    u = x_2^{\alpha_1} (T_{3,1} T_{3,2} T_{4,1} T_{4,2}T_{6,1} T_{6,2})^{\alpha_2}\,,
\end{equation}
with $\alpha_{1,2}$ defined in eq.~(\ref{eq:alphai_massive}) and $T_{i,j}$ with $i \in \{3,4,6\}$ and $j \in \{1,2\}$ defined in eq.~(\ref{eq:Ti_massive}).\\
The integral family we investigate\footnote{We omit any prefactor.} is therefore (with $\mathcal{C}=(0,\infty) \times (1, \infty)^3$)
\begin{equation}
    \int_{\mathcal{C}} u \frac{ dx_2 \wedge dt_3 \wedge dt_4  \wedge  dt_6}{x^{a_2}_2\,T^{a_7}_{3,1}T^{a_8}_{3,2} T^{a_9}_{4,1}T^{a_{10}}_{4,2} T^{a_{13}}_{6,1}T^{a_{14}}_{6,2}\, S_1^{a_{15}}}\,, \quad a_i \in  \mathbb{Z}\,, 
\end{equation}
and
\begin{equation}
    S_1=k_3 t_3{+}k_4 t_4{+}t_6 Y{+}x_2
\end{equation}
Given $X^\vee=\mathbb{C}^4 \setminus \{x_2{=}0, t_3 {\pm} 1{=}0,t_4 {\pm} 1{=}0, t_6 {\pm} 1{=}0\}$, we have one non zero dual relative cohomology, restricted on $X^{\vee} \cap S_1$. Critical point analysis shows this cohomology is $8$ dimensional. 

We give our choice for the basis elements, as well as the corresponding system of differential equations derived with the algorithm of Subsection~\ref{subsec:DEQfromIBPs}, in the ancillary files associated to this work.
\section{Solving the differential equations for internally massive correlator with conformally coupled external states } \label{sec4}
In this Section we focus on the solution of the system of differential equations for the correlator with externally conformally coupled states, derived in eq.~(\ref{eq:def_eq_A}). In order to do this, we introduce the rescaled variables 
\begin{equation}
    \hat{k}=\frac{k_{12}}{k_{34}}\,, \qquad \hat{Y}=\frac{Y}{k_{34}}\,.
\end{equation}
Since $k_{34}$ is the only dimensionful variable it can be set to one, and eventually  recovered through dimensional analysis.
The differential equations are linear in both the parameters $\alpha_1$ and $\alpha_2$, 
\begin{align}
\alpha_1=\frac{1}{2}-\nu+\epsilon \, , \quad \alpha_2=\nu-\frac{1}{2}   \, , \label{eq:alph12_def}
\end{align}
where $\nu=\sqrt{\frac{1}{4}-m^2}$\,.\\
~\\
The limit $m^2 \to 0$, implying $\nu \to 1/2$, gives the conformally coupled case, and was already discussed in eq.~(\ref{eq:limit_Hankel}).  We opt for solving the differential equation as an expansion for small values of the mass. We consider the parameter $\delta$, defined as
\begin{align}
    \delta  =\frac{1}{2}-\nu \overset{m^2 \to 0}{=} m^2 + \mathcal{O}(m^4)\,.
\label{eq:delta_definition}
\end{align}
From eq.~(\ref{eq:delta_definition}) it is clear that an expansion in $\delta$ captures corrections due to a small mass $m^2$. We will therefore organize the calculation in powers of $\delta$, and consider the relations
\begin{align}
    \alpha_1=\delta+\epsilon\,, \quad \alpha_2=-\delta\,. \label{eq:alpha_def}
\end{align}
At this point it is worth stressing that we have two parameters $\epsilon$ and $\delta$ in our integrals.\\ 
We consider an expansion of the form
\begin{equation}
    \mathbf{I}( \epsilon, \delta,\hat{k},\hat{Y}) = \sum_{i= \epsilon_{\text{min}}}^{\infty} \epsilon^i \mathbf{I}_{i} (\delta, \hat{k},\hat{Y})\,.
    \label{eq:series}
\end{equation}
Upon replacing $\alpha_1$ and $\alpha_2$ according to eq.~(\ref{eq:alpha_def}), it is clear that the matrix $\mathbf{A}$ in eq.~(\ref{eq:def_eq_A}) is linear in $\epsilon$ and $\delta$; therefore we can write 
\begin{equation}
\begin{aligned}
   \mathbf{A}( \alpha,\hat{k},\hat{Y})= \mathbf{A}( \epsilon, \delta,\hat{k},\hat{Y})&=\delta A_{\delta}(\hat{k},\hat{Y})+\epsilon A_{\epsilon}(\hat{k},\hat{Y})\,. \label{eq:A_decom_ep_del}\\
    \end{aligned}
\end{equation}
In order to obtain the expression for the master integrals, we start by \textit{formally} solving eq.~(\ref{eq:A_decom_ep_del}) in terms of a path-ordered exponential
\begin{equation}
\begin{aligned}
 \mathbf{I}( \delta,  \epsilon, \hat{k},\hat{Y})&=\mathbb{P} \exp\Big[ \int_\gamma \mathbf{A} \Big] \mathbf{I}( \delta,  \epsilon,\hat{k}_0,\hat{Y}_0)\,, 
    \label{eq:deqA}
\end{aligned}
\end{equation}
where $\gamma$ is a path connecting the boundary point $(\hat{k}_{0},\hat{Y}_0)$ to $(\hat{k},\hat{Y})$. Expanding the path--ordered exponential, and rearranging the expansion in powers of $\epsilon$, we arrive at the following 
\begin{equation}
        \mathbf{I}( \delta,  \epsilon, \mathbf{k},Y) =\left( \epsilon^0 F_0(\delta,\hat{k},\hat{Y})+\epsilon  F_1(\delta,\hat{k},\hat{Y})+\epsilon^2  F_2(\delta,\hat{k},\hat{Y})+...\right)\cdot \mathbf{I}( \delta,  \epsilon,\hat{k}_0,\hat{Y}_0)\,.\label{eq:expansion_epsilon}
\end{equation}
Each of the $F_i(\delta,\mathbf{k},Y)$ can be organized as a series in powers of $\delta$; the first few terms read
\begin{equation}
    \begin{aligned}        
        F_0(\delta,\mathbf{k},Y)&=\mathbf{1}+ \delta\int_\gamma A_{\delta}+ \delta^2 \int_\gamma A_{\delta}\cdot A_{\delta}+ \mathcal{O}(\delta^3)\,, \\
        F_1(\delta,\mathbf{k},Y)&= \int_\gamma A_\epsilon+ \delta \left(\int_\gamma A_{\delta} \cdot  A_{\epsilon} +  \int_\gamma A_{\epsilon} \cdot   A_{\delta}\right)+ \mathcal{O}(\delta^2)\,,  \\
        F_2(\delta,\mathbf{k},Y)&= \int_\gamma A_\epsilon \cdot   A_\epsilon+\delta \Bigg(\int_\gamma A_{\delta} \cdot  A_{\epsilon} \cdot  A_{\epsilon}  + \int_\gamma A_{\epsilon} \cdot  A_{\epsilon} \cdot  A_{\delta}\\
        &\hspace{3.5cm} +\int_\gamma A_{\epsilon} \cdot  A_{\delta} \cdot  A_{\epsilon}\Bigg)+ \mathcal{O}(\delta^2)\,. 
    \end{aligned}
\label{eq:expansionDEQ}
\end{equation}
We should notice that the expansions of the matrices $F_i(\delta,\hat{k},\hat{Y})$ are meaningful since $\delta$ can be assumed to be a small parameter, and truncating the expansion implies neglecting $\delta$-suppressed contributions.

In the next Section we explain how we use the Mellin-Barnes transformation to obtain the asymptotic behaviour of our master integrals. We can then extract the boundary constants $\mathbf{I}( \delta,  \epsilon,\hat{k}_0,\hat{Y}_0)$ from those asymptotic behaviour. These boundary constants have the following expansion in $\epsilon$ 
\begin{equation}
    \mathbf{I}( \delta,  \epsilon,\hat{k}_0,\hat{Y}_0)=\sum_{i= -1}^{\infty}\epsilon^i \, \mathbf{I}_{i}(\delta,\hat{k}_0,\hat{Y}_0).
    \label{eq:BC_generic}
\end{equation}
Since we introduced $\epsilon$ as a regulator, we are interested in the finite $\epsilon$ to zero limit. Given the fact that the coefficients of the decomposition eq.~(\ref{eq:deco_RC}) do not exhibit poles in $\epsilon$, and taking into account eq.~(\ref{eq:BC_generic}), we need only to consider $F_0(\delta,\hat{k},\hat{Y})$ and $F_1(\delta,\hat{k},\hat{Y})$ terms in eq.~(\ref{eq:expansion_epsilon}). Therefore we can safely consider the expansion
{\small
\begin{equation}
  \begin{aligned}
    \mathbf{I}( \epsilon,\delta,\hat{k},\hat{Y}) =  & \frac{1}{\epsilon}F_{0}(\delta,\hat{k},\hat{Y}) \cdot \mathbf{I}_{-1}(\delta,\hat{k}_0,\hat{Y}_0)+F_0(\delta,\hat{k},\hat{Y}) \cdot \mathbf{I}_0(\delta,\hat{k}_0,\hat{Y}_0) \\
    & + F_1(\delta,\hat{k},\hat{Y})\cdot \mathbf{I}_{-1}(\delta,\hat{k}_0,\hat{Y}_0)+\mathcal{O}(\epsilon)\,,
    \end{aligned} \label{eq:PO_eps_expansion}
\end{equation}}
which is sufficient for our purposes.
\subsection{Boundary conditions}

As stated above in order to solve the system of differential equations for the basis introduced in eq.~(\ref{eq:def_eq_A}), we choose the boundary constants at the point $(\hat{Y},\hat{k})=(0,0)$. Since the basis elements are logarithmically divergent at $\hat{Y}=0$, we need to extract their asymptotic behavior. As an illustrative example, let us first investigate the integral associated to $\phi_1$
\begin{equation}
\begin{aligned}
\text{I}(\phi_1)=&\int_{0}^{\infty} \prod_{i=1}^2 d x_i  \int_{1}^{\infty} \prod_{j=1}^2 d t_j \frac{\left(t_1-1\right){}^{\alpha _2-1} \left(t_1+1\right){}^{\alpha _2} \left(t_2-1\right){}^{\alpha _2-1} \left(t_2+1\right){}^{\alpha _2} x_1^{\alpha _1-1} x_2^{\alpha _1}}{\hat{k}+1-t_1 \hat{Y}+t_2 \hat{Y}+x_1+x_2} 
\\
=&\hat{Y}^{-4 \alpha _2} \int_{0}^{\infty} \prod_{i=1}^2  d x_i  \int_{0}^{\infty} \prod_{j=1}^2 d \tilde{t}_j \frac{\tilde{t}_1^{\alpha _2-1} \tilde{t}_2^{\alpha _2-1} x_1^{\alpha _1-1} x_2^{\alpha _1} \left(\tilde{t}_1+2 \hat{Y}\right){}^{\alpha _2} \left(\tilde{t}_2+2 \hat{Y}\right){}^{\alpha _2}}{\hat{k}+1-\tilde{t}_1+\tilde{t}_2+x_1+x_2} \, . \label{intbc}
\end{aligned}
\end{equation}
For later convenience, we considered the change of variable $t_i \to \tilde{t}_i/\hat{Y}+1$. The factor $\hat{Y}^{-4\alpha_2}$  corresponds to part of the asymptotic behavior of $\text{I}(\phi_1)$, leading to logarithmic divergences in the limit $\hat{Y}\to0$. 

The integral in eq.~(\ref{intbc}) is non-trivial to perform; focusing for concreteness on the  $\tilde{t}_1-$integration, we notice that there are three hyperplanes involving this variable, namely $\tilde{t}_1$, $\tilde{t}_1+2\hat{Y}$ and $\hat{k}+1-\tilde{t}_1+ \tilde{t}_2+x_1+x_2$ (the same holds true considering the $\tilde{t}_2-$integration). In order to simplify the computation, we introduce the Mellin-Barnes transformation
\begin{align}
(\tilde{t}_i+2\hat{Y})^\alpha= \frac{1}{2 \pi i\, \Gamma (-\alpha )} \int_{\gamma_i-i\infty}^{\gamma_i+i \infty} dz_i \Gamma \left(-z_i\right) \Gamma\left(z_i-\alpha \right) \frac{(2 \hat{Y})^{z_i} }{ \tilde{t}_i ^{z_i-\alpha }}\,, \qquad i =1,2\,,
\label{eq:Mellin-Barnes}
\end{align}
where $\gamma_i$ is a path separating the poles of $\Gamma(-z_i)$ from the ones of $\Gamma(z_i-\alpha)$.
Thanks to eq.~(\ref{eq:Mellin-Barnes}), the number of hyperplanes involving the variable $\tilde{t}_1$ is reduced to two (once again the same holds true for $\tilde{t}_2$) and the integration can be performed in terms of Euler's Gamma functions. Therefore eq.~(\ref{intbc}) reduces to
\begin{equation}
\begin{aligned}
 \text{I}(\phi_1)=& -\frac{\Gamma \left(  \delta+\epsilon\right) \Gamma \left(\delta+\epsilon+1\right) }{\pi^2   \Gamma \left(\delta\right){}^2} (\hat{k}+1)^{2 \delta+2\epsilon} \prod_{i=1}^2 \int_{\gamma_i-i\infty}^{\gamma_i+i \infty} dz_i   2^{-2+z_1+z_2} e^{-i \pi  \left(z_1+2 \delta\right)}
\\
&  \left(\frac{\hat{Y}}{\hat{k}+1}\right)^{4 \delta+z_1+z_2} \Gamma \left(-z_1\right) \Gamma \left(-z_2\right) \Gamma \left(-2 \delta-z_1\right) \Gamma \left(-2 \delta-z_2\right)  \Gamma \left(z_1+\delta\right) 
\\
& \Gamma \left(z_2+\delta\right) \Gamma \left(z_1+z_2+2 \delta-2\epsilon\right) \, ,
\end{aligned} \label{bdcon}
\end{equation}
where we have replaced $\alpha_1$ and $\alpha_2$ with eq.~(\ref{eq:alpha_def}). In order to perform the remaining integration in the Mellin-Barnes variables $z_i$, we are free to add to eq.~(\ref{bdcon}) the arc at plus infinity obtaining a contour integral. The latter reduces to a double sum over residues evaluated at $z_i=n$ and $z_i=n-2\delta$, where $n \in \mathbb{N} \cup 0$. However, since the residues for $n>0$ are power suppressed in $\hat{Y}$, only the residues at $z_i=0$ and $z_i=-2\delta$ are relevant; taking this into account we obtain\footnote{The direction in which we approach the point $(\hat{k}, \hat{Y})=(0,0)$ is, in principle, a delicate issue e.g. due to the presence of multiple letters vanishing at the origin. However, restricting ourself to the situation in which the invariants are positive, the Mellin-Barnes representation shows that no logarithmically divergent factor depending on $(\hat{k}-\hat{Y})$, and therefore sensitive to the order of the limits, is produced.}
\begin{equation}
\begin{aligned}
\lim_{(\hat{Y},\hat{k}) \to (0,0)} \text{I}(\phi_1)=&  \frac{ 2^{-4 \delta}  \Gamma \left(  \delta+\epsilon\right) \Gamma \left(\delta+\epsilon+1\right) }{ \Gamma \left(\delta\right){}^2} \Big[ \Gamma \left(-2 (\delta+\epsilon)\right) \Gamma \left(2 \delta\right){}^2 \Gamma \left(-\delta\right){}^2
\\
&+2^{2 \delta} \Gamma \left(2 \delta\right) \Gamma \left(\delta\right) \Gamma \left(-\delta\right) \Gamma \left(-2\delta\right) \Gamma \left(-2 \epsilon\right) \hat{Y}^{2 \delta} 
\\
&+ 2^{4 \delta} e^{-2 i \pi  \delta} \Gamma \left(\delta\right){}^2 \Gamma \left(-2 \delta\right){}^2 \Gamma \left(-2 \left(-\delta+\epsilon\right)\right) \hat{Y}^{4 \delta} 
\\
&+ 2^{2 \delta} e^{-2 i \pi  \delta} \Gamma \left(2 \delta\right) \Gamma \left(\delta\right) \Gamma \left(-\delta\right) \Gamma \left( -2\delta\right) \Gamma \left(-2 \epsilon\right) \hat{Y}^{2 \delta}\Big] \, .
\end{aligned}
\label{eq:Mellin_Barnse_Gamma}
\end{equation}
It is worth noticing that the factor $\Gamma \left(-2 \epsilon\right)$ gives a pole in $\epsilon$; this reflects the fact that, without introducing this parameter as regulator, this integral would be otherwise divergent. Therefore we can Laurent expand the constant part of eq.~(\ref{eq:Mellin_Barnse_Gamma}) in $\epsilon$ obtaining
{\small
\begin{equation}
\begin{aligned}
    \lim_{(\hat{Y},\hat{k}) \to (0,0)} \text{I}(\phi_1)&=-\frac{  2^{-2 \delta}\left(1+e^{-2 i \pi  \delta}\right)
   \Gamma (-2 \delta) \Gamma (-\delta) \Gamma (2 \delta) \Gamma (\delta+1)}{2\epsilon} \, +\mathcal{O}(\epsilon^0)\,.
    \end{aligned}
    \label{eq:bd_expansion}
\end{equation}
}
For the remaining integrals, the boundary constants can be computed in a similar fashion. We notice that for the last four integrals, both the hyperplanes $S_1$ and $S_2$ appear in the integrand and therefore it is necessary to introduce additional Mellin-Barnes transformations.

\begin{figure}
     \centering
     \includegraphics[scale=.5]{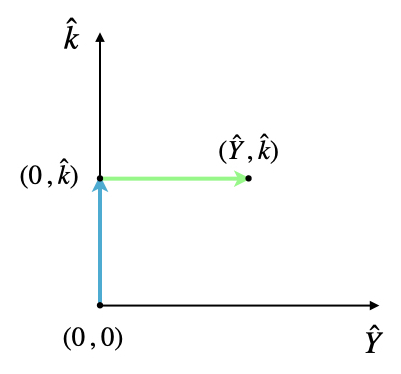}
     \caption{Explicit integration path $\gamma$ chosen for the solution eq.~(\ref{eq:deqA}); the integration along the blu line is performed as first step, followed by the integration along the green line.}
     \label{fig:pathgamma}
 \end{figure}

Having obtained the boundary constants at the point $(\hat{Y}, \hat{k})=(0,0)$, we integrate eq.~(\ref{eq:expansionDEQ}) along the path $\gamma$ depicted in Figure \ref{fig:pathgamma}: first from the origin $(0,0)$ to $(0,\hat{k})$, and then from this point to a generic $(\hat{Y},\hat{k})$ which we choose in the region
\begin{equation}
\hat{k}>\hat{Y}>0 \quad \cap \quad 0 < \hat{Y}<1\,.
\label{eq:region_validity_solution}
\end{equation} 
Given the fact that all the entries in the matrix $ \mathbf{A}( \delta,\epsilon,\hat{k},\hat{Y})$ are in $d\log$ form (whose arguments are linear functions in the kinematic variables), the series solution eq.~(\ref{eq:series}) can be expressed in terms of multiple polylogarithms~\cite{GoncharovPaper}. These functions are defined recursively as
\begin{equation}
\begin{split}
    G(a_1, \dots , a_n ; z) & = \int_0^z \frac{dt}{t-a_1} \, G(a_2, \dots , a_n ; t)\,, \qquad a_i \in \mathbb{C}\,,\\ \qquad G(\underbrace{0, \dots , 0}_{\text{$n$ times}};z) & \equiv \frac{1}{n!} \log^n (z)\, .
\end{split}
\end{equation}
We have now all the ingredients to evaluate explicitly eq.~(\ref{eq:PO_eps_expansion}), namely all the master integrals up to finite order in $\epsilon$, and therefore also eq.~(\ref{eq:deco_RC}). As mentioned several times the corresponding expression for $\text{I}(\varphi_{LC'})$ can be obtained upon the replacement $k_{12} \leftrightharpoons k_{34}$. We notice that under this transformation, the region\footnote{Once expressed in terms of the original variables $k_{12}, k_{34}$ and $Y$.} in eq.~(\ref{eq:region_validity_solution}) is mapped into itself. Therefore, as far as we limit ourselves to the above-mentioned region, we do not have to consider possible complications related to the analytic continuations of multiple polylogarithms appearing in the result.

\subsection{Analytic result}
After these considerations we are able to present the analytic expansion of the correlator eq.~(\ref{eq:correlator_complete}) in terms of a small mass parameter $\delta$. We stress now that while, on the one hand, individual master integrals exhibit poles in the regulator $\epsilon$, they all cancel in the correlator eq.~(\ref{eq:correlator_complete}) as alluded in Subsection~\ref{sec:correlatot_def}. Let us consider
\begin{equation}
    \begin{aligned}
         \psi_2(x,y)&=\frac{\lambda^2_{1,2}}{2 Y}\sum_{i=0}^\infty  \psi^{(i)}_2(x,y) \delta^i, \\
               \end{aligned}
\end{equation}
where we introduced the variables $x=\frac{Y-k_{12}}{Y+k_{12}}$ and $y=\frac{Y-k_{34}}{Y+k_{34}}$ for convenience. The first few coefficients of the expansion can be written as\footnote{We simplified the expressions with \textsc{PolyLogTools}~\cite{Duhr:2019tlz}.}
\begin{align}
 \psi^{(0)}_2(x,y)=&G(0,1,x y)-G(0,-1,x)-G(0,-1,y)-\frac{\pi ^2}{6} \, ,
 \label{eq:masslesslimit}
 \\
 \psi^{(1)}_2(x,y) = &-G(-1,x) G(0,-1,y)-\frac{1}{6} \pi ^2 G(-1,x)-G(-1,y) G(0,-1,x) -\frac{1}{6} \pi ^2 G(-1,y)
  \nonumber \\
 &-G(-1,0,-1,x)+G(0,-1,-1,x)-G(-1,0,-1,y)+G(0,-1,-1,y) 
\nonumber \\
 &-G(0,0,1,x y)+G(-x,0,1,x y)+G(-y,0,1,x y) -3 \zeta (3) \, .
 \label{eq:firstcorrection}
\end{align}
Some comments are in order. On the one hand,  eq.~(\ref{eq:masslesslimit}) corresponds to the conformally coupled limit and, upon matching the conventions\footnote{This amounts to $k_{12} \to X_1$ and $k_{34} \to X_2$; furthermore we have an over-all factor $(-1)$ due to the imaginary unit assigned to each coupling in eq.~(\ref{eq:correlator_ext_massless}).}, it reproduces the result in \cite{Hillman:2019wgh}, see also~\cite{Arkani-Hamed:2015bza,De:2023xue}. On the other hand, eq.~(\ref{eq:firstcorrection}) is our prediction for the first order mass correction to the tree-level two-site correlator with internal massive states. We have reported here the first two orders in the expansion. In principle higher order terms in the expansion can be obtained algorithmically.

\section{On the singularities of cosmological correlators}
\label{sec5}
In this Section we investigate the origin of the singularities appearing in the differential equations. Specifically we find that all these singularities or, equivalently, the symbol letters can be expressed as Pl\"ucker coordinates associated with hyperplanes in the integrand.  

Let us consider an integral of the form
\begin{align}
\Psi= \int_{\mathcal{C}} d^{n+1}\mathbf{Z} \prod_i^m (\mathbf{X}_i \cdot \mathbf{Z})^{\gamma_i}  \,,
\label{eq:projectiv_int}
\end{align}
where the exponents $\gamma_i$ may also be integer, and the integration variables are promoted to projective space and collectively denoted by $\mathbf{Z}$. For instance, for the correlator eq.~(\ref{eq:psi_RC}) we have $\mathbf{Z}=\{z_0, \dots ,z_4\}$ and, in the chart $z_0=1$ we have the identification $\{z_1, \dots,z_4\}=\{ x_1, x_2, t_1,t_2 \}$. The $\mathbf{X}_i$ is the normal vector to the $i-$th hyperplane appearing in the integrand. We can group all the vectors into a matrix as (the $i-$th column is given by $\mathbf{X}_i$)
\begin{align}
\mathcal{A}=\left( \mathbf{X}_1 \, \mathbf{X}_2, \cdots \, \mathbf{X}_m \right) \, ,
\end{align}
and the Pl\"ucker coordinates are defined as $\langle \mathbf{X}_{i_1} \dots \mathbf{X}_{i_{n+1}} \rangle \equiv \text{det}(\mathbf{X}_{i_1} \dots \mathbf{X}_{i_{n+1}})$. 

\begin{figure}[hbt]
\centering
\includegraphics[width=0.6\textwidth]{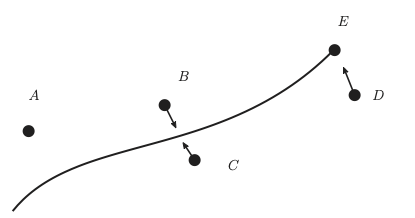}
\caption{Different types of singularities in the integrand. Point A is an isolated singularity, which can be avoided by contour deformation. Point E is an ``end point singularity"; points D is pinched with E. Points B and C are examples of ``pinched singularities".}
\label{fig:sigularities}
\end{figure}

In a one dimensional integral, there is an intuitive description linking the singularities of the integrated results in terms of the singularities of the integrand. The poles of the integrand are functions of external kinematic variables; therefore, changing the values of the kinematic variables will move the position of the poles of the integrand. In general, if one of the singularities occurs along the integration path, it is possible to deform the latter away from the above-mentioned singularity. However, there are two special cases: one situation, referred to as ``pinched singularity", is when two singular points are pinching the contour from both sides. The other case, known as ``end point singularity", is when the singular points are approaching the end points of the contour (the situation is depicted in Figure~\ref{fig:sigularities}). In these two cases, the corresponding kinematic value, i.e. the configuration of the kinematics such that one of the two special cases occur, is a potential singular point of the integral. 
Here we propose to link pinched and end-point singularities to a kinematic configuration such that a collection of the vectors, say $\mathbf{X}_{i_1} \dots \mathbf{X}_{i_{n+1}}$ becomes linearly dependent. Equivalently, the point determined by $\langle \mathbf{X}_{i_1}, \dots, \mathbf{X}_{i_{n+1}} \rangle=0$ is a potential singular point of the integrals eq.~(\ref{eq:projectiv_int}).

Let us consider a two-dimensional simple example, whose hyperplanes are $z_1, z_2$ and $-a+z_1+z_2$ and $a$ is a kinematic variable. The corresponding vectors $\mathbf{X}_i$ are
\begin{equation}
\begin{aligned}
    \mathbf{X}_1&=\{0,1,0 \} \, , \\
    \mathbf{X}_2&=\{0,0,1\} \, , \\
    \mathbf{X}_3&=\{-a, 1,1 \} \, .
\end{aligned}
\end{equation}
\begin{figure}[hbt]
\centering
\includegraphics[width=0.55\textwidth]{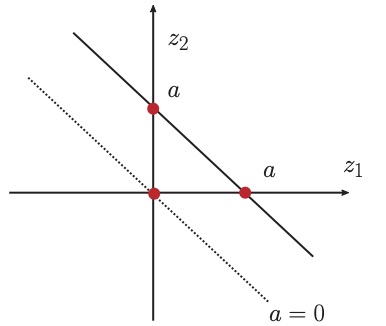}
\caption{Pinching of the intersection points in the limit $a \to 0$.}
\label{fig:example}
\end{figure}
The poles are along hyperplane $\{z_1=0, z_2=0, z_1+z_2-a=0\} $ as well as the intersection points $(0,0)$, $(0,a)$ and $(a,0)$. When $a\to 0$, all these three intersection points will be pinched at $(0,0)$, as shown in Figure~\ref{fig:example}. Therefore, the integral is singular at $a=0$. 
At this point, the vectors $\mathbf{X}_1$, $\mathbf{X}_2$ and $\mathbf{X}_3$ are no longer linearly independent. Equivalently, $\langle \mathbf{X}_{1} \mathbf{X}_2 \mathbf{X}_3 \rangle \to 0$ indicates a singular point of the integral \footnote{Recently, there was a development in the understanding of Landau singularities from a topological perspective, see \cite{Fevola:2023kaw,Fevola:2023fzn}. In these works the authors proposed that the so called Landau singularities for Feynman integrals are associated to drop of certain  Euler characteristics, that is to say the number of closed chambers enclosed by hyperplanes will reduce at the given kinematic configuration. The above consideration concerning the degeneracy of the vectors $\mathbf{X}_{i_1}, \dots, \mathbf{X}_{i_{n+1}}$ can be understood in a similar way. Since $\langle \mathbf{X}_{i_1} \dots \mathbf{X}_{i_{n+1}} \rangle \to 0$, the corresponding closed chamber will collapse to a lower dimensional object.
}.

Now, let us turn to investigates the singularities of integral in eq.~(\ref{eq:psi_RC}). The matrix $\mathcal{A}$ is given by 
\begin{align}
\mathcal{A}=\left(
\begin{array}{cccccccc}
 0 & 0 & -1 & 1 & -1 & 1 & k_{34} & k_{12}+k_{34} \\
 1 & 0 & 0 & 0 & 0 & 0 & 0 & 1 \\
 0 & 1 & 0 & 0 & 0 & 0 & 1 & 1 \\
 0 & 0 & 1 & 1 & 0 & 0 & 0 & -Y \\
 0 & 0 & 0 & 0 & 1 & 1 & Y & Y \\
\end{array}
\right) \, , 
\end{align}
where the projective variables are defined as $\mathbf{Z}=\{1,x_1,x_2,t_1,t_2 \}$. By computing all the Pl\"ucker coordinates, one would reproduce the letters as given by a direct calculation in eq.~(\ref{eq:letter}). Similarly, for the fully massive case, the matrix $\mathcal{A}$ reads
\begin{align}
\mathcal{A}=\left(
\begin{array}{cccccccccccccccc}
 0 & 0 & -1 & 1 & -1 & 1 & -1 & 1 & -1 & 1 & -1 & 1 & -1 & 1 & 0 & 0 \\
 1 & 0 & 0 & 0 & 0 & 0 & 0 & 0 & 0 & 0 & 0 & 0 & 0 & 0 & 0 & 1 \\
 0 & 1 & 0 & 0 & 0 & 0 & 0 & 0 & 0 & 0 & 0 & 0 & 0 & 0 & 1 & 1 \\
 0 & 0 & 1 & 1 & 0 & 0 & 0 & 0 & 0 & 0 & 0 & 0 & 0 & 0 & 0 & k_1 \\
 0 & 0 & 0 & 0 & 1 & 1 & 0 & 0 & 0 & 0 & 0 & 0 & 0 & 0 & 0 & k_2 \\
 0 & 0 & 0 & 0 & 0 & 0 & 1 & 1 & 0 & 0 & 0 & 0 & 0 & 0 & k_3 & k_3 \\
 0 & 0 & 0 & 0 & 0 & 0 & 0 & 0 & 1 & 1 & 0 & 0 & 0 & 0 & k_4 & k_4 \\
 0 & 0 & 0 & 0 & 0 & 0 & 0 & 0 & 0 & 0 & 1 & 1 & 0 & 0 & 0 & -Y \\
 0 & 0 & 0 & 0 & 0 & 0 & 0 & 0 & 0 & 0 & 0 & 0 & 1 & 1 & Y & Y \\
\end{array}
\right) \, .
\label{eq:Amatrix_fully_massive}
\end{align}
There are 40 letters appearing in the differential equations. By comparing with the Pulcker coordinates obtained from eq.~(\ref{eq:Amatrix_fully_massive}) we obtain the following letters
\begin{equation}
\begin{aligned}
\{ &k_1,k_2,k_3,k_4,Y,k_1+k_2+k_3+k_4,k_1+k_2+k_3-k_4, k_1+k_2-k_3-k_4, 
\\
& -k_1-k_2+k_3+k_4,k_1-k_2+k_3+k_4, k_1+k_2-k_3+k_4,k_1-k_2+k_3-k_4, 
\\
&k_1-k_2-k_3+k_4,-k_1+k_2+k_3-k_4, -k_1+k_2+k_3+k_4, -k_1+k_2-k_3+k_4,
\\
& k_1+k_2+k_3+k_4-2 Y, k_1+k_2+k_3+k_4+2 Y, -k_1-k_2-k_3+k_4+2 Y,
\\
&  k_1+k_2+k_3-k_4+2 Y,-k_1-k_2+k_3-k_4+2 Y,k_1+k_2-k_3+k_4+2 Y,
\\
&-k_1-k_2+k_3+k_4+2 Y,k_1+k_2-k_3-k_4+2 Y,-k_1+k_2-k_3-k_4+2 Y,
\\
& k_1-k_2+k_3+k_4+2 Y,-k_1+k_2-k_3+k_4+2 Y, k_1-k_2+k_3-k_4+2 Y,
\\
&-k_1+k_2+k_3-k_4+2 Y,k_1-k_2-k_3+k_4+2 Y,
 -k_1+k_2+k_3+k_4+2 Y,
 \\
 &k_1-k_2-k_3-k_4+2 Y, -k_1-k_2+Y,-k_1+k_2+Y,k_1-k_2+Y,
 \\
 &k_1+k_2+Y,k_3-k_4+Y,-k_3+k_4+Y,k_3+k_4+Y,-k_3-k_4+Y \} \, .
\end{aligned}
\end{equation}

\section{Conclusions}
In this paper we considered two site tree-level cosmological correlators with massive propagator in de Sitter universe. Given the integration over conformal time, as well as the presence of Hankel functions in the integrand, the naive integral representation for the correlator is non trivial to evaluate. Utilizing relations among Hankel functions and Bessel functions, the integral representation of the latter, as well as suitably chosen Fourier transformations for certain factors depending on conformal time in the integrand, we derived a four-fold integral representation for the correlator. This representation is written as a sum of three contributions dubbed as right-center, left-center and left-right. Each of those terms can be written in terms of hyperplanes; furthermore the integrands admit a natural decomposition in terms of a multi-valued part and a rational (holomorphic) differential form. The structure is therefore similar to the one of Feynman integrals appearing in multi-loop calculus (in dimensional regularisation), and it was possible to borrow techniques often used in this field, such as integration by parts, reduction to master integrals and the method of differential equations, to obtain analytic expressions for the correlator.    

In particular, in Section \ref{sec3} we discussed the two site tree-level cosmological correlator with massive internal state and conformally coupled external states; we focused on the so called right-center contribution to the correlator (given the fact that the left-center contribution can be obtained from this by a trivial replacement and the left-right can be integrated directly). We embedded the individual integral into a more general integral family and studied the latter using tools from the framework of twisted cohomology, or, more precisely, of \emph{relative} twisted cohomology. This is the proper mathematical framework to tackle integrals in which some of the singularities present in the differential forms are not ``regulated" (they are not present) in the multivalued part of the integral. It was possible to identify a basis of eight master integrals and derive a system of first order differential equations for the basis elements with respect to each of the kinematic variables appearing in the problem. As a consistency check, we derived differential equations using standard integration by parts identities.  Our set up was also applied to the integral families related to the fully massive case (i.e. including massive external states) as described in Subsection \ref{fullM}. In this case twisted cohomology predicted a basis of $128$ master integrals; we provide our basis choice, as well as the corresponding differential equations (obtained with the algorithm of Subsection~\ref{subsec:DEQfromIBPs}) in an ancillary file.

In Section \ref{sec4} the solution of differential equations associated to the internally massive externally conformally coupled correlator was obtained as an expansion with respect to a parameter, denoted by $\delta$, which can be thought of as a small mass (we therefore present the solution as a ``small mass expansion"). On a technical level the boundary vector for the system of differential equations was obtained via the Mellin-Barnes representation and the solutions was written, order by order in $\delta$ in terms of multiple polylogarithms.

We verified that the massless limit of our result, corresponding to the constant term in the $\delta$ expansion of the solution, is in agreement with the known result in the literature. This offers a consistency check of our method and set-up. The expansion in the parameter $\delta$ could be, in principle, obtained up to the desired order with no conceptual obstacle.

In Section \ref{sec5} we made some remarks on the singularities of the integral, borrowing ideas and intuition from the study of Landau singularities in the context of Feynman integrals. A natural question consists in relating singularities of the integral to singularities of the integrand; the latter are known as ``pinched" or ``end point" singularities. For the case cosmological correlators considered in this work, we observed that both type of singularities can be related to Pl\"ucker coordinate of the matrix associated to the hyperplanes configuration. 
\section*{Acknowledgements}
We are indebted to Stefan Weinzierl for several pieces of advice on different aspects of this work; we would like to thank Andrzej Pokraka for a discussion concerning relative twisted cohomology. We are grateful to Paolo Benincasa, Hayden Lee, Pierpaolo Mastrolia, Andrzej Pokraka and Stefan Weinzierl for comments and feedback on an earlier version of this draft.\\
This work has been supported by the Cluster of Excellence Precision Physics,
Fundamental Interactions, and Structure of Matter (PRISMA EXC 2118/1) funded by the German Research Foundation (DFG) within the German Excellence Strategy (Project ID 390831469). Xiaofeng Xu is supported by the funding from the European Research Council (ERC) under the
European Union’s Horizon 2022 Research and Innovation Program (ERC Advanced Grant
agreement No.101097780, EFT4jets). Views and opinions expressed are however those of the author(s) only and do not necessarily reflect those of the European Union or
the European Research Council Executive Agency. Neither the European Union nor the granting authority can be held responsible for them.
\newpage
\appendix

\section{Algebra of operators}
\label{app:algebra_operators}
The goal of this Appendix is to clarify the action of the operator $\nabla^{\vee}$ and $\nabla^{\vee}_{\text{kin}}$ on relative dual forms; for further discussions see~\cite{Caron-Huot:2021iev,Caron-Huot:2021xqj,De:2023xue}. First let us introduce the action of the differential operator $d(\bullet)$ on $\theta = \prod_i^\ell \theta_i$; we have
\begin{equation}
    d \theta = \sum_j \delta_j \theta\,;
    \label{eq:dtheta}
\end{equation}

Let us denote with $K=(i_1, \dots, i_k)$ a generic multi-index; let us consider an element of the form
\begin{equation}
   \varphi^{\vee} = \sum_{K} \delta_K( \theta \varphi_K^{\vee})\,.
\label{eq:dual_form}
\end{equation}
Treating formally $\delta_K$ as a $K$-differential form and using eq.~(\ref{eq:dtheta}) we get
\begin{equation}
    \begin{aligned}
         \nabla^\vee (\delta_K (\theta \varphi^{\vee}_K))&=(d+\omega^\vee \wedge) (\delta_K (\theta \varphi^{\vee}_K))= (-1)^{k}(\delta_K (\theta (d+\omega_K \wedge)\varphi_K))+(-1)^{k}(\delta_K (d\theta  \varphi_K)) \\
        &=(-1)^{k}\delta_K ( \theta\,\nabla^\vee_{K} \varphi_K)+(-1)^{k}\sum_{m \notin K} \delta_{Km} (\theta \,\varphi^{\vee}_{Km})\,,  \label{eq:delta_proof}
    \end{aligned}
\end{equation}
where $\omega_K$ (resp. $\nabla^{\vee}_K$) stands for $\omega$ (resp. $\nabla^{\vee}$) restricted on  $X^{\vee} \cap_{j=1}^k S_{i_j}$, and (once again formally) $\delta_{Km} = \delta_K \wedge \delta_{m}$.\\
~\\
Let us now comment on the action of $\nabla^{\vee}_{\text{kin}}$ on dual forms. In order to build intuition, we can think at an alternative to eq.~(\ref{eq:nabla_kin_op}) valid for ``dual integrals", i.e. eq.~(\ref{eq:hyperint}) with the replacements $u \to u^{-1}$, $\varphi \to \varphi^{\vee}$ and $\mathcal{C} \to \mathcal{C}^{\vee}$. Using eq.~(\ref{eq:nabla_kin_op}) and localizing the integral on $X^{\vee} \cap_{j=1}^k S_{i_j}$ as an effect of $\delta_K$ 
\begin{equation}
    I^{\vee}( \varphi^{\vee}) = \int_{\mathcal{C}^{\vee}} u^{-1} \varphi^\vee = \sum_K \int_{\mathcal{C}^{\vee}_{K}} u^{-1}_{K} (\theta \varphi^{\vee}_K)\,,
\end{equation}
where, again, $u^{-1}_K$ (resp. $\mathcal{C}^{\vee}_{K}$) stands for $u^{-1}$ (resp. $\mathcal{C}^{\vee}$) restricted on  $X^{\vee} \cap_{j=1}^k S_{i_j}$. Proceeding as in eq.~(\ref{eq:nabla_kin_op}) we obtain
\begin{equation}
    d_{\text{kin}}I^{\vee}( \varphi^{\vee}) = \sum_K 
    \left(
    \int_{\mathcal{C}^{\vee}_{K}} u^{-1}_{K} (\theta \,  \nabla^{\vee}_{\text{kin}}\big|_{K}\,\varphi^{\vee}_K)+\sum_{m \notin K} \int_{\mathcal{C}^{\vee}_{K}} u^{-1}_{K} (\theta \,d_{\text{kin}}(\theta_m) \wedge\varphi^{\vee}_K)
    \right)\,,
\label{eq:nabla_kin_dual_integral}
\end{equation}
where the last term in eq.~(\ref{eq:nabla_kin_dual_integral}) can further restrict the integrand on $X^{\vee} \cap_{j=1}^k S_{i_j} \cap S_m$. Finally from eq.~(\ref{eq:nabla_kin_dual_integral}) we infer
\begin{equation}
 \nabla^{\vee}_{\text{kin}} ( \delta_K ( \theta \varphi^{\vee}_K)) = \delta_K( \theta \nabla^{\vee}_{\text{kin}}\big|_{K} \varphi^{\vee}_K) + \sum_{m \notin K}\delta_{Km} ( \theta d_{\text{kin}} (\theta_m) \wedge \varphi^{\vee}_{Km} )\,.
\end{equation}
\clearpage

\bibliographystyle{JHEP}
\bibliography{bibcf.bib}

\end{document}